\documentclass[a4paper,fleqn,usenatbib]{mnras}
\usepackage{float}
\usepackage[T1]{fontenc}
\usepackage{ae,aecompl}
\usepackage{graphicx}
\usepackage{amsmath}
\usepackage{amssymb}
\usepackage{supertabular}
\usepackage{siunitx}

\def\beq{\begin{eqnarray}}
\def\eeq{\end{eqnarray}}

\title[Neutrinos and GWs from magnetized NDAFs]
{Neutrinos and gravitational waves from magnetized neutrino-dominated accretion discs with magnetic coupling}

\author[Song, Liu \& Wei]
{Cui-Ying Song$^{1,2}$, Tong Liu$^{1}$\thanks{E-mail:tongliu@xmu.edu.cn}, Yun-Feng Wei$^{1}$ \\
$^{1}$ Department of Astronomy, Xiamen University, Xiamen, Fujian 361005, China\\
$^{2}$ School of Physics and Astronomy, Monash University, Melbourne, Victoria 3800, Australia}

\date{Accepted XXX. Received YYY; in original form ZZZ}

\pubyear{2020}

\begin{document}
\label{firstpage}
\pagerange{\pageref{firstpage}--\pageref{lastpage}}
\maketitle

\begin{abstract}
Gamma-ray bursts (GRBs) might be powered by a black hole (BH) hyperaccretion systems via the Blandford-Znajek (BZ) mechanism or neutrino annihilation from neutrino-dominated accretion flows (NDAFs). Magnetic coupling (MC) between the inner disc and BH can transfer angular momentum and energy from the fast-rotating BH to the disc. The neutrino luminosity and neutrino annihilation luminosity are both efficiently enhanced by the MC process. In this paper, we study the structure, luminosity, MeV neutrinos, and gravitational waves (GWs) of magnetized NDAFs (MNDAFs) under the assumption that both the BZ and MC mechanisms are present. The results indict that the BZ mechanism will compete with the neutrino annihilation luminosity to trigger jets under the different partitions of the two magnetic mechanisms. The typical neutrino luminosity and annihilation luminosity of MNDAFs are definitely higher than those of NDAFs. The typical peak energy of neutrino spectra of MNDAFs is higher than that of NDAFs, but similar to those of core-collapse supernovae. Moreover, if the MC process is dominant, then the GWs originating from the anisotropic neutrino emission will be stronger particularly for discs with high accretion rates.
\end{abstract}


\section{Introduction}\label{sec:intro}

It is well known that a hyperaccreting stellar-mass black hole (BH) surrounded by a disc is one of the leading central engine models of gamma-ray bursts \citep[GRBs, e.g.,][]{Narayan1992,Woosley1993,MacFadyen1999,Popham1999,Ruffert1999}. The Blandford-Znajek \citep[BZ,][]{Blandford1977} process and neutrino annihilation are two possible mechanisms to power GRB jets. In the BZ process, the BH ergosphere and the remote astrophysical load are connected by the open magnetic field lines supported by external currents flowing in an equatorial disc, and an electric potential difference will be induced. If the field strength is large enough, the vacuum is unstable to a cascade production of electron-positron pairs, and a surrounding force-free magnetosphere will be established. Then, the rotating energy and angular momentum of the BH can be extracted and transported to the remote load under these circumstances \citep[e.g.,][]{Blandford1977,Lee2000a,Lee2000b}. For high accretion rates, in the inner regions of the hyperaccretion disc, the temperature and density are so high that photons are trapped. The disc can only be cooled by neutrino emission. For this reason, these discs are named neutrino-dominated accretion flows \citep[NDAFs, see][for reviews]{Liu2017a,Liu2018,Liu2019}. Annihilation of some fraction of the emitted neutrinos produces a relativistic electron-positron outflow to power a GRB \citep[e.g.,][]{Ruffert1997,Asano2000,Zalamea2011}. The ability of NDAFs to power GRBs has been studied by many authors \citep[e.g.,][]{Narayan2001,Kohri2002,Di2002,Gu2006,Chen2007,Kawanaka2007,Liu2007,Lei2009,Lei2017,Song2015,Song2016,Yi2017a}.

The strong gravitational waves (GWs) from BH hyperaccretion systems released by anisotropic neutrino emission \citep[e.g.,][]{Sago2004,Suwa2009,Liu2017b} or precession of NDAFs \citep[e.g.,][]{Reynoso2006,Lei2007,Romero2010,Liu2010,Sun2012} have been widely discussed. \citet{Sun2012} found that DECIGO and BBO might detect GWs from BH-inner-disc precession systems at $\sim10~\rm Hz$, particularly for GRBs in the Local Group ($\lesssim 1 ~\rm Mpc$). \citet{Liu2017b} summarized and compared the GWs from three GRB central engine models, i.e., NDAFs, the BZ mechanism, and millisecond magnetars. The BZ mechanism cannot release GWs once it coexists with the neutrino annihilation in NDAFs, and the GW power is entirely determined by the neutrino luminosity. In the collapsar scenarios, the typical frequency of GWs triggered by the anisotropic neutrino emission is $\sim 1-100 \rm Hz$, and the masses and metallicities of the progenitor stars have slight effects on the GW strains \citep[e.g.,][]{Wei2020}.

As one of the variants of the BZ process, the magnetic coupling (MC) process has been proposed by some authors \citep[e.g.,][]{Blandford1999,Li2000,Li2002}. In this process, the field lines are closed and directly connect the central BH and the disc; thus, the energy and angular momentum are not taken away to a remote load but are instead exchanged between the BH and the accretion disc. The effects of MC torques have been investigated in some disc models. For example, \citet{Li2002} found that the MC process may significantly change the local radiative flux in thin discs, \citet{Wang2002} and \citet{Wang2003a} discussed a condition for the coexistence of the BZ and MC processes on thin discs, and \citet{Ma2007} investigated the effects of the MC process on advection-dominated accretion flows. More recently, the effects of MC torque on NDAFs have been investigated \citep[e.g.,][]{Lei2009,Luo2013}. The results indicted that the neutrino luminosity and neutrino annihilation luminosity can both be efficiently enhanced by MC. However, it is scarcely mentioned that the BZ and MC mechanisms coexist in NDAFs.

If the BZ and MC mechanisms coexist in NDAFs, then the effects on the structure and radiation of NDAFs are worth exploring. Since the MC torque can enhance the neutrino luminosity, the competition between the two mechanisms will inevitably lead to a change in the GWs powered by the anisotropic neutrino emission. In this paper, we investigate magnetized NDAFs (MNDAFs) assuming the coexist of the BZ and MC processes. This paper is organized as follows. In Section 2, we describe the MNDAF model and the methods to calculate its neutrino and GW emission. The effects of magnetic fields on the neutrino, neutrino annihilation, and GW emission are presented in Section 3. The conclusions and discussion are given in Section 4.

\section{Model}

Short GRBs (SGRBs) might originate from the mergers of neutron star (NS)-NS or NS-BH binaries, and long GRBs (LGRBs) are related to the massive collapsars. In both scenarios, a stellar-mass BH surrounded by an NDAF might be formed. The BZ mechanism can extract the energy and angular momentum from the BH through the large-scale magnetic fields. Moreover, the MC process also can transfer energy and angular momentum between the BH and the disc. Here we consider BZ and MC mechanism coexist in the BH accretion system.

The BZ power and torque are given by \cite{Wang2002} in the context of an equivalent circuit,
\beq
P_{\rm BZ}=2 P_{0} a_{*}^{2}\int_{0}^{\theta _{\rm c}}\frac{\kappa(1-\kappa)\sin^{3}\theta d \theta }{2-(1-q)\sin^{2}\theta},
\eeq
\beq
T_{\rm BZ}=4 T_{0} a_{*}(1+q)\int_{0}^{\theta _{\rm c}}\frac{(1-\kappa)\sin^{3}\theta d \theta }{2-(1-q)\sin^{2}\theta},
\eeq
where $\kappa \equiv \Omega _{\rm F}/\Omega _{\rm H}$ denotes the ratio of the angular velocity of the remote astrophysical load to that of the horizon, $\Omega _{\rm H}=a_{*}c^{3}/\left [ 2 G M_{\rm BH}(1+q) \right]$, and $q=a_{*}/(1+\sqrt{1-a_{*}^{2}})$, where $a_{*}\equiv c J_{\rm BH}/GM_{\rm BH}^{2}$ is the dimensionless spin parameter of the BH,  $M_{\rm BH}$ and $J_{\rm BH}$ are the mass and angular momentum of the BH. Without any valuable knowledge of the remote load, the value of $\kappa$ is uncertain. The optimal BZ power can be obtained for $\kappa$ = 0.5 \citep[e.g.,][]{MacDonald1982}. We set $\kappa$ = 0.2 and 0.5 in the calculations in order to present its effects on the structure of the magnetic fields.

Moreover, $P_{0}\approx 6.59\times 10 ^{50}B_{\rm H, 15}^{2} m_{\rm BH}^{2}$, $T_{0}\approx 3.26\times 10 ^{45}B_{\rm H, 15}^{2} m_{\rm BH}^{3}$, $B_{\rm H, 15}=B_{\rm H}/(10^{15} ~\rm G)$, and $m_{\rm BH}=M_{\rm BH}/M_\odot$. The magnetic field strength at the horizon is $B_{\rm H}= \sqrt{8 \pi \dot{M} c}/r_{\rm g}$, where $\dot{M}$ is the accretion rate (the dimensionless accretion rate $\dot{m}$ = $\dot{M}/M_\odot~\rm s^{-1}$) and the gravitational radius $r_{\rm g}=G M_{\rm BH}/c^{2}$ \citep[e.g.,][]{McKinney2005}.

The magnetic field in the inner parts of the disc is complicated and currently not well understood \citep[see a detailed study by][]{Jafari2018}, and many authors have recently researched this topic \citep[see e.g.,][]{Lovelace2009,Bisnovatyi2007,Bisnovatyi2012}. To simplify the calculations, we assumed that the magnetic fields in the disc vary as $ B_{\rm D} \propto \xi ^{-n}$,  as described in \citet{Blandford1976}. The dimensionless radial parameter $\xi $ equals $r/r_{\rm ms}$, and $n$ is the power-law index denoting the degree of the concentration of magnetic fields in the equatorial plane of the disc. Then, we have
\beq
B_{\rm D}=B_{\rm H}\frac{r_{\rm H}}{\varpi _{D}(r_{\rm ms})} \xi ^{-n},
\eeq
where $\varpi _{D}(r_{\rm ms}) = r _{\rm ms}\sqrt{1+ \chi_{\rm ms}^{-4}a_{*}^{2}+ 2\chi_{\rm ms}^{-6}a_{*}^{2} }$,  $\chi _{\rm ms}=(r_{\rm ms}/r_{\rm g})^{1/2}$, the marginally stable orbit radius of the disc $r_{\rm ms}=r_{\rm g}[3+Z_{2}-\sqrt{(3-Z_{1})(3+Z_{1}+2Z_{2})}]$, $Z_{1}=1+(1-a_{*}^{2})^{1/3}[(1+a_{*})^{1/3}+(1-a_{*})^{1/3}]$, and $Z_{2}=\sqrt{3a_{*}^{2}+Z_{1}^{2}}$ for $0 < a_{*} < 1$ \citep[e.g.,][]{Bardeen1972,Kato2008}. Figure 1 shows the radius profiles of the magnetic field strength on the disc $B_{\rm D}$ for the different values of the BH spin $a_{*}$, accretion rate $\dot{m}$, and power-law index $n$.

\begin{figure}
\centering
\includegraphics[width=8cm,height=8cm]{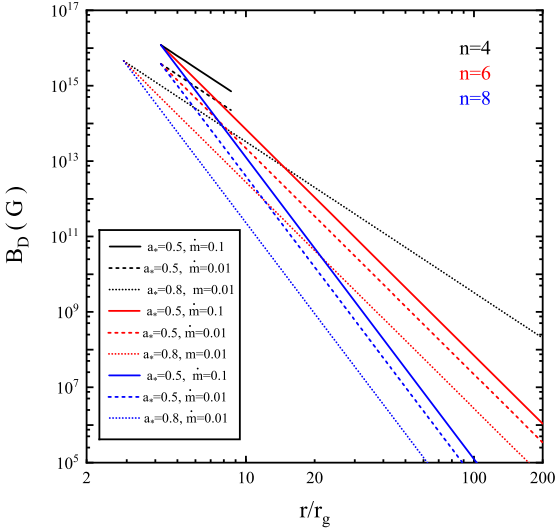}
\caption{The radius profiles of the magnetic field strength on the disc $B_{\rm D}$ for the different values of the BH spin $a_{*}$, accretion rate $\dot{m}$, and power-law index $n$.}
\label{fig1}
\end{figure}

Similarly, we can obtain the total MC power and torque \citep[e.g.,][]{Wang2002,Wang2003a},
\beq
P_{\rm MC}=2 P_{0} a_{*}^{2}\int_{\theta _{\rm c}}^{\frac{ \pi }{2} }\frac{\beta (1-\beta)\sin^{3}\theta d \theta }{2-(1-q)\sin^{2}\theta},
\eeq
\beq
T_{\rm MC}=4 T_{0} a_{*}(1+q)\int_{\theta _{\rm c}}^{\frac{\pi}{2}}\frac{(1- \beta )\sin^{3}\theta d \theta }{2-(1-q)\sin^{2}\theta},
\eeq
where $\beta =\Omega _{\rm D}/\Omega _{\rm H}$ is the ratio of the angular velocity of the disc $\Omega_{\rm D}=[(r^{3}/GM)^{1/2}+a_{*}GM/c^{3}]^{-1}$ to that of the horizon and $\theta_{\rm c}$ indicates the angular boundary between the open and closed field lines, with values equal to 0 (or $\pi/2$) when the BZ (or MC) effect vanishes. Based on conservation of magnetic flux, the mapping relation between the angular coordinate $\theta_{\rm c}$ and the radial coordinate on the disc can be derived as follows \citep[e.g.,][]{Wang2002,Wang2003a}:
\beq
&&\cos \theta _{\rm c}=\int_{1}^{\xi _{\rm out}} d \xi\nonumber \\&&\frac{\xi^{1-n}\chi _{\rm ms}^{2}\sqrt{1+a_{*}^{2}\chi _{\rm ms}^{-4}\xi ^{-2}+2a_{*}^{2}\chi _{\rm ms}^{-6}\xi ^{-3}} }{2\sqrt{(1+a_{*}^{2}\chi _{\rm ms}^{-4}+2a_{*}^{2}\chi _{\rm ms}^{-6})(1-2\chi _{\rm ms}^{-2}\xi ^{-1}+a_{*}^{2}\chi _{\rm ms}^{-4}\xi ^{-2})}},\nonumber \\
\eeq
where $\xi _{\rm out}=r_{\rm out}/r_{\rm ms}$ is the dimensionless outer boundary of the MC region.

The hydrodynamics of the hyperaccretion disc is well approximated by a steady and axisymmetric accretion flow. The coefficient of kinematic viscosity in the usual form as $\nu= \alpha c_{\rm s} H$, where $\alpha$ is a dimensionless parameter that absorbs all the detailed microphysics of the viscous processes, $c_{\rm s}=\sqrt{P/\rho}$ is isothermal sound speed with $P$ and $\rho$, the total pressure and mass density, and $H$ is the half thickness of the disc \citep{Kato2008}.

The basic equations of the MNDAF model are given as follows. Conservation of mass implies that the accretion rate is
\beq
\dot{M}=-4\pi r v_{r}  \rho H,
\eeq
where $v_{r}$ is the radial velocity of the gas.

The equation of angular momentum conservation can be written as
\beq
\dot{M}r^{2}\sqrt{\frac{GM_{\rm BH}}{r^{3}}}\frac{D}{A}+ T_{\rm MC}=4\pi r^{2}H \alpha P\sqrt{\frac{A}{BC}},
\eeq
where $A$, $B$, $C$, and $D$ in the above equations are the well-known relativistic correction factors for a thin accretion disc around a Kerr BH \citep[e.g.,][]{Riffert1995,Liu2010}, which have been widely used in theoretical calculations or simulations of accretion discs. The corrected expression of the disc's half thickness due to hydrostatic equilibrium in the vertical direction $H\simeq \sqrt{{P r^{3}}/{\rho GM_{\rm BH}}}\cdot\sqrt{{B}/{C}}$.

The total pressure $P$ is the sum of the contributions from five terms, including the radiation pressure, the gas pressure, the electron degeneracy pressure, the neutrino pressure, and the magnetic pressure \citep[e.g.,][]{Kohri2005,Gu2006},
\beq
P&=&\frac{11}{12}aT^{4}+\frac{\rho k_{\rm B} T}{m_{\rm p}}(\frac{1+3X_{\rm nuc}}{4})\nonumber \\&&+\frac{2\pi hc}{3}(\frac{3}{8\pi m_{\rm p}})^{4/3}(\frac{\rho }{\mu _{\rm e}})^{4/3}+\frac{u_{\rm  \nu }}{3}+P_{\rm mag},
\eeq
where $X_{\rm nuc}\simeq\min(1, ~295.5 \rho _{10}^{-3/4}T_{11}^{9/8} \exp(-0.8209/T_{11}))$ is the mass fraction of free nucleons \citep[e.g.,][]{Qian1996,Liu2007}, $u_{\nu}$ is the neutrino energy density \citep[for details, see, e.g.,][]{Kohri2005}, $T_{11}=T/(10^{11} \rm~K)$, and $\rho_{10}=\rho/(10^{10} \rm ~g~ {cm}^{-3}) $. Here, $P_{\rm mag}=\beta_{t}P$ is the magnetic pressure contributed by the tangled magnetic field in the disc, and $\beta_{t}$ is the ratio of the magnetic pressure to the total pressure.

\begin{figure*}
\centering
\includegraphics[width=5cm,height=5cm]{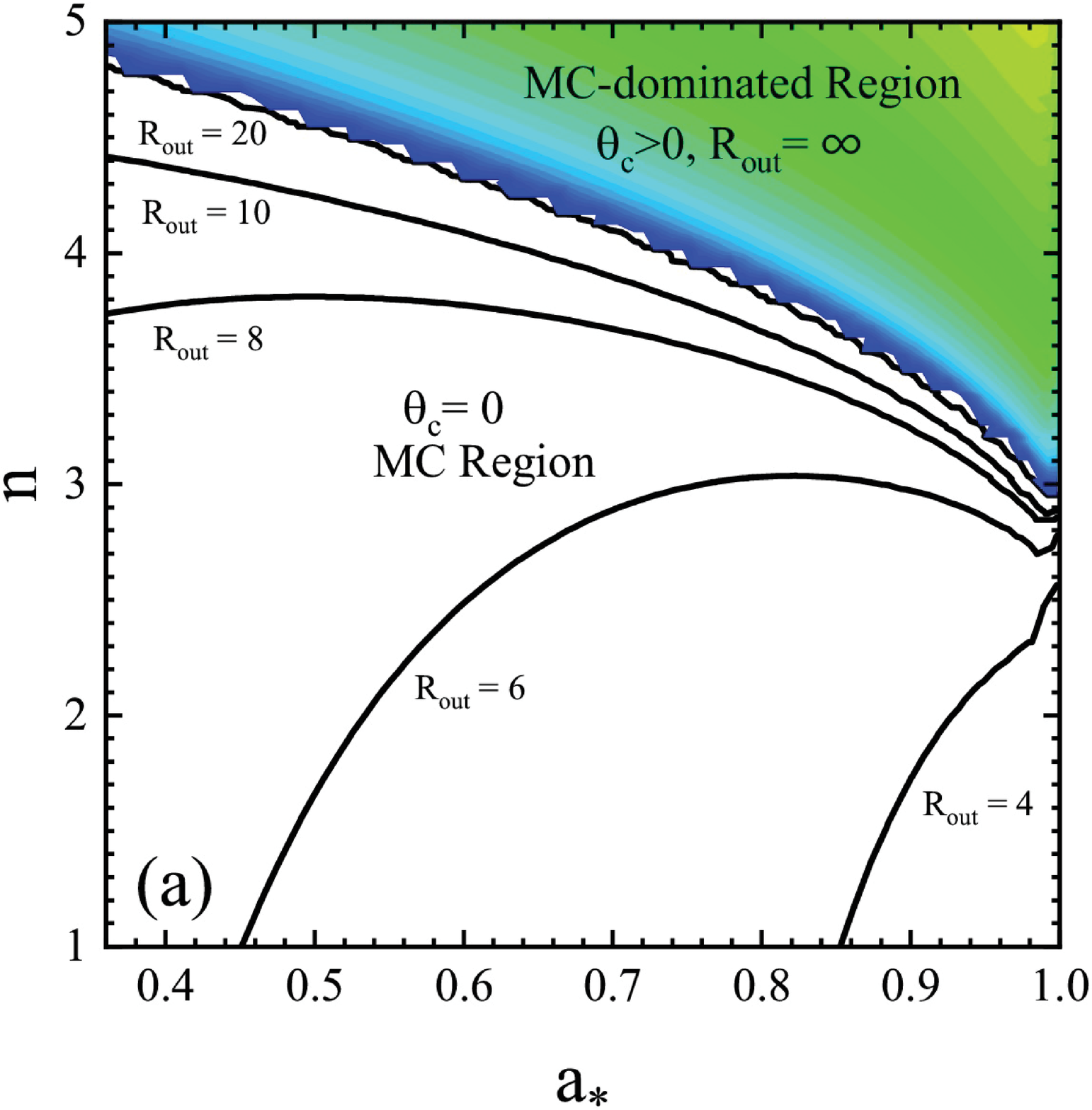}
\includegraphics[width=5cm,height=5cm]{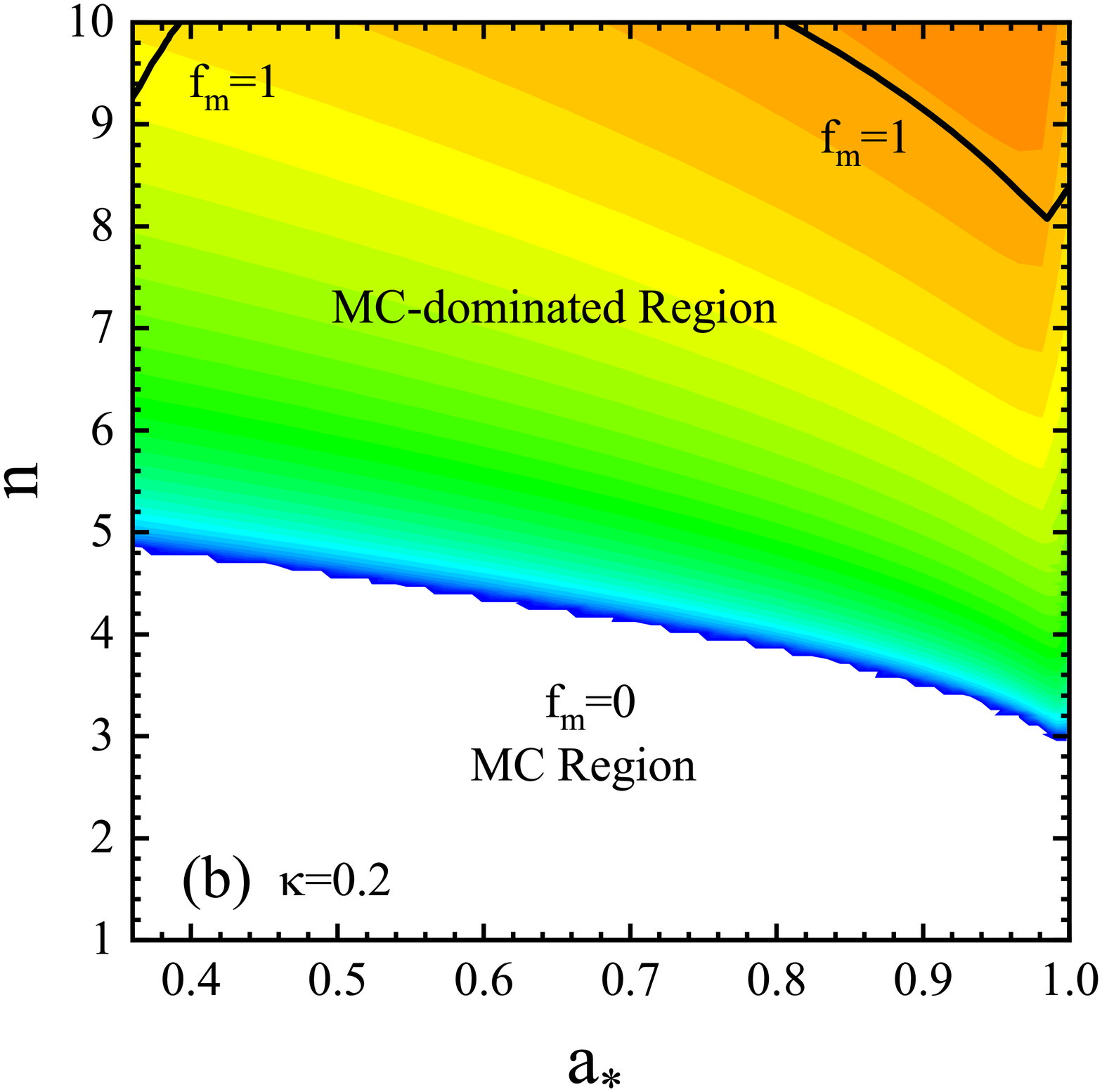}
\includegraphics[width=6cm,height=5cm]{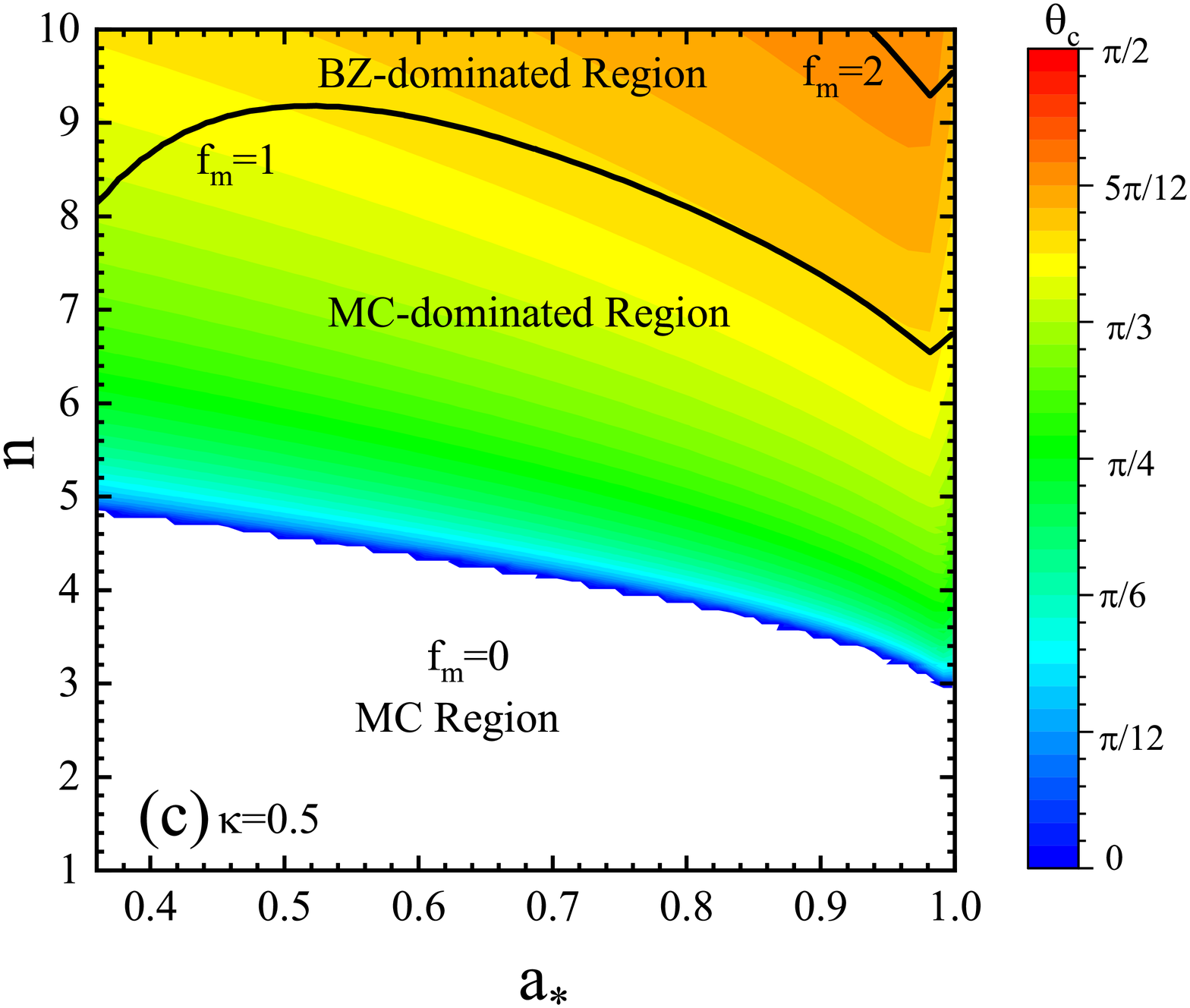}
\caption{(a) The outer boundary of the MC region for the different values of the BH spin $a_{*}$ and power-law index of the magnetic field $n$ under the mapping relation constraint. (b) The values of $\theta_{\rm c}$ and $f_{\rm m}=P_{\rm BZ}/P_{\rm MC}$ vary with $a_{*}$ and $n$ for given $R_{\rm out}=200$, $\kappa=0.2$. (c) Similar to (b) except for $\kappa=0.5$. The values of $\theta_{\rm c}$ from $0$ to $\pi/2$ are indicated by the different colors.}
\label{fig2}
\end{figure*}

Considering conservation of energy, we can obtain the energy balance equation,
\beq
Q_{\rm vis}= Q_{\rm G}^{+} +  Q_{\rm MC}^{+}=Q_{ \rm \nu }^{-}+ Q_{\rm adv}^{-},
\eeq
where $Q_{\rm vis}$ represents the viscous heating rate, including the contributions of the gravitational potential $Q_{\rm G}={3GM\dot{M}}D/{8\pi r^{3}B}$ and the MC effect $Q_{\rm MC}={T_{\rm MC}}/{4\pi r}\cdot{d\Omega_{\rm D}}/{dr}$ \citep[e.g.,][]{Lei2009,Luo2013}. The cooling rates due to neutrino losses $Q_{\rm \nu}$ and advection $Q_{\rm adv}$ have known expressions \citep[e.g.,][]{Kohri2005,Gu2006,Liu2007}.
In the calculations, the radiative cooling $Q_{\rm rad}^{-}$ is negligible compared with the other cooling terms. Moreover, we ignore the cooling rate arising from photodisintegration $Q_{\rm photo}$ because it is much less than the neutrino cooling rate in the inner disc \citep[e.g.,][]{Janiuk2004,Liu2007}.

The neutrino luminosity from MNDAFs is
\beq
L_{\rm \nu }=4\pi \int_{r_{\rm ms}}^{r_{\rm out}}Q_{\rm  \nu }r dr.
\eeq
In this paper, we only consider accretion rate $\dot{M}<1 ~ M_{\odot}~\rm s^{-1}$. In this case, the neutrino trapping effects can be ignored \citep[e.g.,][]{Xue2013}. The neutrino annihilation luminosity is obtained by integrating over the whole space outside the BH and the disc, following the approach in \citet{Ruffert1997}, \citet{Popham1999}, and \citet{Rosswog2003}.

Neutrinos are mainly emitted from the inner region of the disc, so the shapes of the neutrino spectra would be effected by the general relativistic effects. We use the well-known ray-tracing methods \citep[e.g.,][]{Fanton1997,Li2005} to calculate the neutrino propagation effects and then obtain the observed neutrino spectra. For each pixel of the observed image, the position of the emitter on the disc can be traced based on the null geodesic equation \citep{Carter1968}. We can calculate the energy shift of a neutrino by investigating the corresponding velocity and gravitational potential of the emission location. Integrating over all the pixels, the energy extension of a particular rest-frame neutrino emission energy can be obtained. The total observed spectrum can be derived by integrating over all the emission energies. The total observed flux can be expressed as \citep[e.g.,][]{Wei2019}
\beq
F_{E_{\rm obs}}=\int_{\rm image} g^{3} I_{E_{\rm em}}d \Omega_{\rm obs},
\eeq
where $E_{\rm obs}$ is the observed neutrino energy, $E_{\rm em}$ is the neutrino emission energy from the local disc, $\Omega_{\rm obs}$ is the solid angle of the disc image to the observer, and $g\equiv E_{\rm obs}/E_{\rm em}$ is the energy shift factor. $I_{\rm em}$ is the local emissivity \citep{Rauch1994}, which can be calculated according to the cooling rate $Q_{\rm \nu}$ (either cooling rate of electron neutrinos $Q_{\rm \nu_{e}}$ or electron anti-neutrinos $Q_{\bar{\nu} _{e}}$) as
\beq
I_{E_{\rm em}}=Q_{\nu }\frac{F_{E_{\rm em}}}{\int F_{E_{\rm em}}d E_{\rm em}},
\eeq
where $F_{E_{\rm em}}=E_{\rm em}^{2}/[\exp (E_{\rm em}/ k_{\rm B} T-\eta )+1)]$ is the unnormalized Fermi-Dirac spectrum.

Moreover, the GWs from a BH hyperaccretion system induced by the anisotropic emission of neutrinos might be detectable \citep[e.g.,][]{Liu2017a}. The general expression for the local GW flux is \citep[e.g.,][]{Suwa2009}
\beq
\frac{d E_{\rm GW}}{dA dt}=\frac{d P_{\rm GW}}{dA}=\frac{c^{3}}{16 \pi G}|\frac{d h_{+}(t)}{dt}|^{2},
\eeq
where $dA=D_{L}^{2}d\Omega_{\rm obs}$ represents the surface element, $D_{\rm L}$ denotes the distance between the observer and the source, and $h_{+}(t)$ is the nonvanishing GW amplitude of NDAFs \citep[e.g.,][]{Mueller1997}.

Integrating over a sphere surrounding the source, we obtain the average GW power,
\beq
P_{\rm GW}=\frac{1}{9} \gamma \frac{G}{c^{5}}L_{\nu}^{2},
\eeq
where $\gamma=43/3-20\ln2 \approx 0.47039$.

Following the calculations of \citet{Suwa2009}, the characteristic GW strain can be expressed as
\beq
h_{c}(f)=\sqrt{\frac{2 G}{{\pi}^2 c^3 D_{L}^2} \frac{dE_{\rm GW}(f)}{df}}.
\eeq

For a single burst event, the above equation can be evaluated by:
\beq
h_{c}(f)=\frac{\sqrt{\gamma}h_{\infty}}{\pi^{2} f{T_{\rm GRB}}}| \sin(\pi {T_{\rm GRB}} f )|,
\eeq
and
\beq
h_{\infty}\sim 1.8\times 10^{-21}(\frac{\rm 10~Mpc}{D_{\rm L}})(\frac{L_{\nu} T_{\rm GRB}}{10^{54} ~\rm ergs}),
\eeq
where $T_{\rm GRB}$ is the activity timescale of the central engine of GRBs.

Consider the case of multiple bursts, one finds the characteristic strain as:
\beq
h_{c}(f)=\frac{\sqrt{\gamma}h_{\infty}}{\pi^{2}N \delta t f}|\frac{\sin (\pi \delta t  f) \sin(\pi {T_{\rm GRB}} f)}{\sin (\pi {T_{\rm GRB}} f/N)}|,
\eeq
where $N$ is the number of sub-bursts and $\delta t$ is the duration of one sub-burst.

Since the GW characteristic strains are given, we can also compute signal-to-noise ratios (SNRs) obtained from matched filtering in GW detectors. The optimal SNR is defined by
\beq
{\rm SNR}^2=\int_{0}^{\infty} d({\rm ln}f)\frac{h_{c}(f)^{2}}{h_{n}(f)^{2}},
\eeq
where $h_{n}(f)=[5f S_{h}(f)]^{1/2}$ is the noise amplitude with $S_{h}(f)$ being the spectral density of the strain noise in the
detector at frequency $f$.

\section{Results}
\subsection{Structure and Luminosities of MNDAFs}

\begin{figure*}
\centering
\includegraphics[width=16cm,height=8cm]{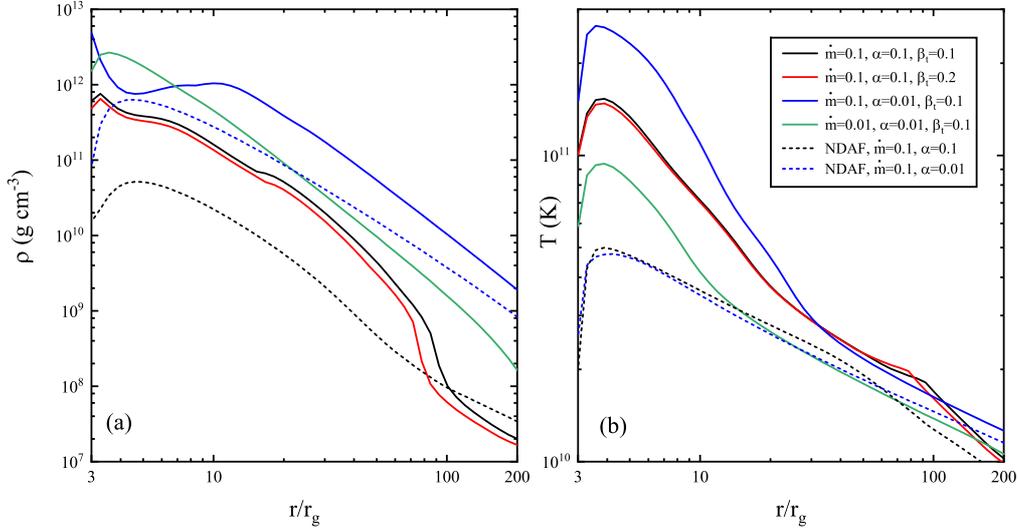}
\caption{Radial profiles of the density $\rho$ and temperature $T$ of MNDAFs and NDAFs for the different accretion rate $\dot{m}$, viscous parameter $\alpha$, and ratio of the magnetic pressure to the total pressure $\beta_{t}$ with $n=4$ and $a_{*}=0.8$.}
\label{fig3}
\end{figure*}

To visualize the impact of different parameters on the MC and BZ processes coexisting in the magnetic field configurations, Figure 2 shows the variation in the magnetic field strength with the power-law index $n$, the BH spin $a_{*}$, and the dimensionless radial parameter $R_{\rm out}=r_{\rm out}/r_g$. Here, we take BH spin $a_{*}>0.36$, in this condition, the energy and angular momentum transfer direction is always from the BH to the disc \citep[e.g.,][]{Wang2003b}.

As shown in Figure 2(a), the parameter space can be divided by the value of the critical angle $\theta_{\rm c}$ into the MC region (uncolored part, $\theta_{\rm c}=0$) and the MC-BZ coexistence region (colored part, $\theta_{\rm c}>0$). In the MC region, the dimensionless outer boundary of the closed magnetic fields $R_{\rm out}$ remains finite. The parameter $R_{\rm out}$ approaches infinity when the MC and BZ processes coexist. If $R_{\rm out}=200$ is fixed, we can calculate the ratio of the BZ and MC powers, $f_{\rm m}=P_{\rm BZ}/P_{\rm MC}$, for the different $n$ and $a_{*}$. As shown in Figures 2(b) and 2(c), the ranges of $f_{\rm m}=0$, $f_{\rm m}<1$, and $f_{\rm m}>1$ correspond to the MC region, the MC-dominated region, and the BZ-dominated region, respectively. Meanwhile, there are various critical angles $\theta_{\rm c}$ in the different regions. The color gradient from red to blue indicates the values of the critical angle $\theta_{\rm c}$ from large to small. It should be noted that the uncertainty in the ratio of the angular velocity of the remote load to that on the horizon, $\kappa$ will influence the value of $P_{\rm BZ}$. Moreover, in order to explore the effect of $\kappa$ on the structure of the magnetic fields, we calculated the BZ power by taking $\kappa$=0.2 and 0.5 in Figures 2(b) and 2(c), respectively. It is easy to find that, for the same $n$ and $a_{*}$, the values of $f_{\rm m}$ in Figure 2(b) is generally smaller than these in Figure 2(c). In other words, $\kappa=0.5$ corresponds to the smaller MC-dominated region in the same parameter space.

Besides the parameters related to the magnetic fields, there are some other parameters belonging to the accretion systems. The radial profiles of the mass density and temperature of the discs with different $\dot{m}$, $\alpha$ and $\beta_{t}$ are shown in Figure 3. In the calculations, we fix $M_{\rm BH}=3M_{\odot}$, $a_{*}=0.8$, and $n=4$. The solid lines and the dashed lines represent the NDAFs with and without the MC process, respectively. The density and temperature of MNDAFs are greatly improved by the MC process. Moreover, $\rho$ and $T$ increase with increasing $\dot{m}$, and decrease with increasing $\alpha$ and $\beta_{t}$.

Figure 4 displays the variations in the neutrino luminosity $L_{\nu}$, the neutrino annihilation luminosity $L_{\nu \bar{\nu}}$, the GW power $P_{\rm GW}$, and the BZ power $P_{\rm BZ}$ with the mass accretion rate $\dot{M}$ for the different $\alpha$ and $\beta_{t}$. Here we also fix $M_{\rm BH}=3M_{\odot}$, $a_{*}=0.8$ and $n=4$. The black, red, and green solid lines represent $L_{\nu}$, $L_{\nu \bar{\nu}}$, and $P_{\rm GW}$, in turn. Moreover, the blue and cerulean lines indict $P_{\rm BZ}$ for $\kappa=0.5$ and $0.2$, respectively. The cases of NDAFs excluding magnetic fields are indicated by colorful dashed lines. In the Figures, it can be noted that the neutrino annihilation luminosity is less than the BZ power when the accretion rate $\dot{M} \lesssim 0.01 ~ M_{\odot}~\rm s^{-1} $. As the accretion rate increases, the neutrino annihilation luminosity gradually exceeds the BZ power and eventually dominates. Comparing 4(a) with 4(b), one can find that $\beta_{t}$ have little effects on the neutrino luminosity, the annihilation luminosity, and the GW power. From Figures 4(a) and 4(c), the luminosities/powers with $\alpha = 0.01$ are larger than those with $\alpha = 0.1$.

We also plot the variations of $L_{\nu}$, $L_{\nu \bar{\nu}}$, $P_{\rm GW}$, and $P_{\rm BZ}$ with $\dot{M}$ for the different $a_*$ and $n$ as shown in Figure 5. Here we fix $\alpha=0.1$, $\beta_{t}=0.1$, and $\kappa=0.5$. Similar to Figure 4, different lines denote different luminosity/power. As defined in Section 2, $\theta_{\rm c}$ indicates the angular boundary between the BZ and MC magnetic fields. Greater $n$ and $a_{*}$ correspond to larger $\theta_{\rm c}$, and thus the BZ magnetic fields are stronger relative to the total magnetic fields. Comparing Figure 5(a) with 5(b), it is obvious that these four luminosities/powers increase significantly with $a_{*}$ for both MNDAFs and NDAFs. In Figure 5(c), the BZ mechanism is always greater than the neutrino annihilation luminosity. Compared to NDAFs without magnetic fields, $L_{\nu}$, $L_{\nu \bar{\nu}}$, and $P_{\rm GW}$ of MNDAFs are greatly enhanced as shown in Figures 4 and 5. The strength of GWs will also vary. The $P_{\rm GW}$ of MNDAFs is several orders of magnitude greater than that of NDAFs with the typical frequency of $\sim$ 10 Hz.

\begin{figure*}
\centering
\includegraphics[width=0.3\textwidth]{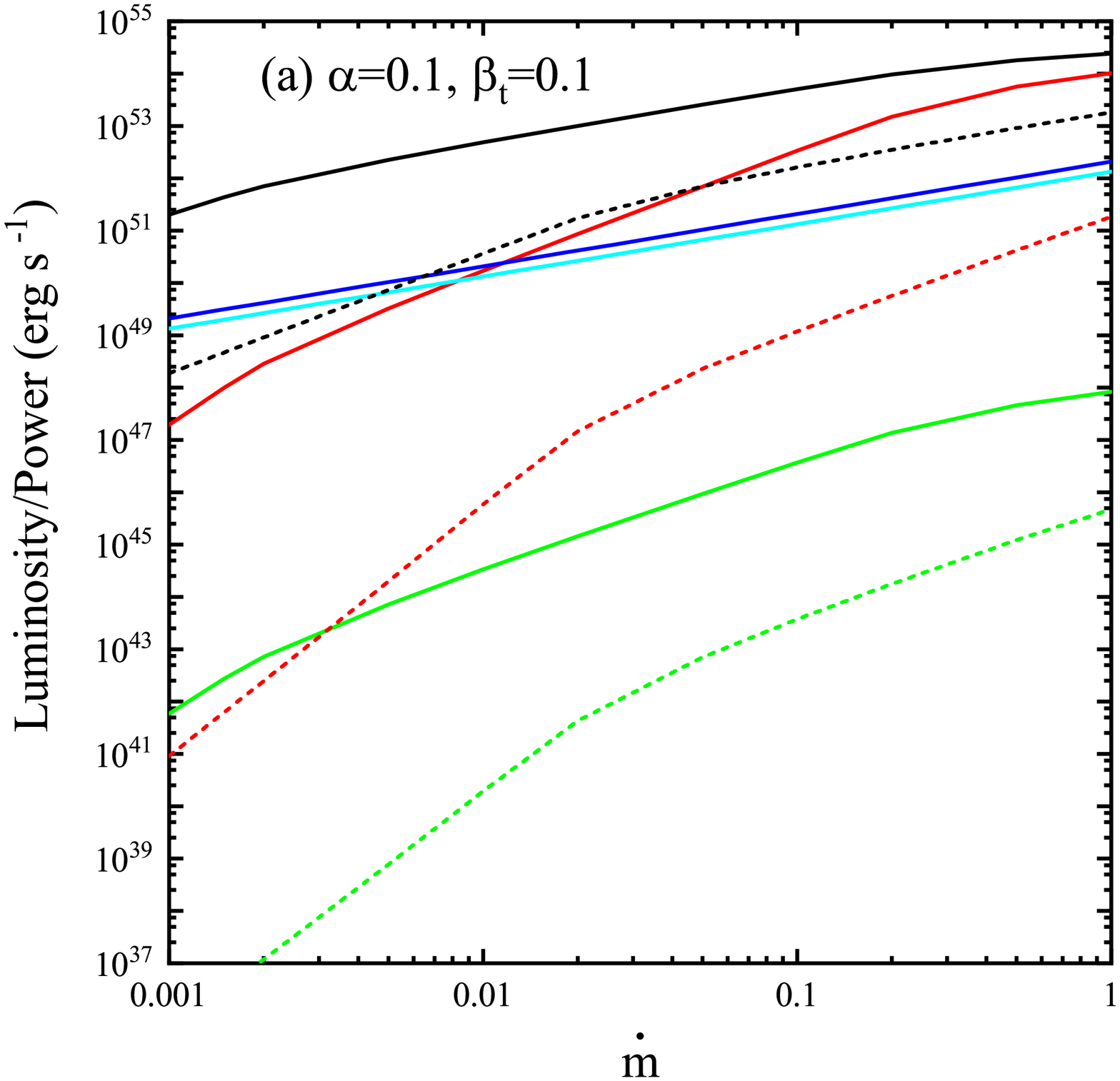}
\includegraphics[width=0.3\textwidth]{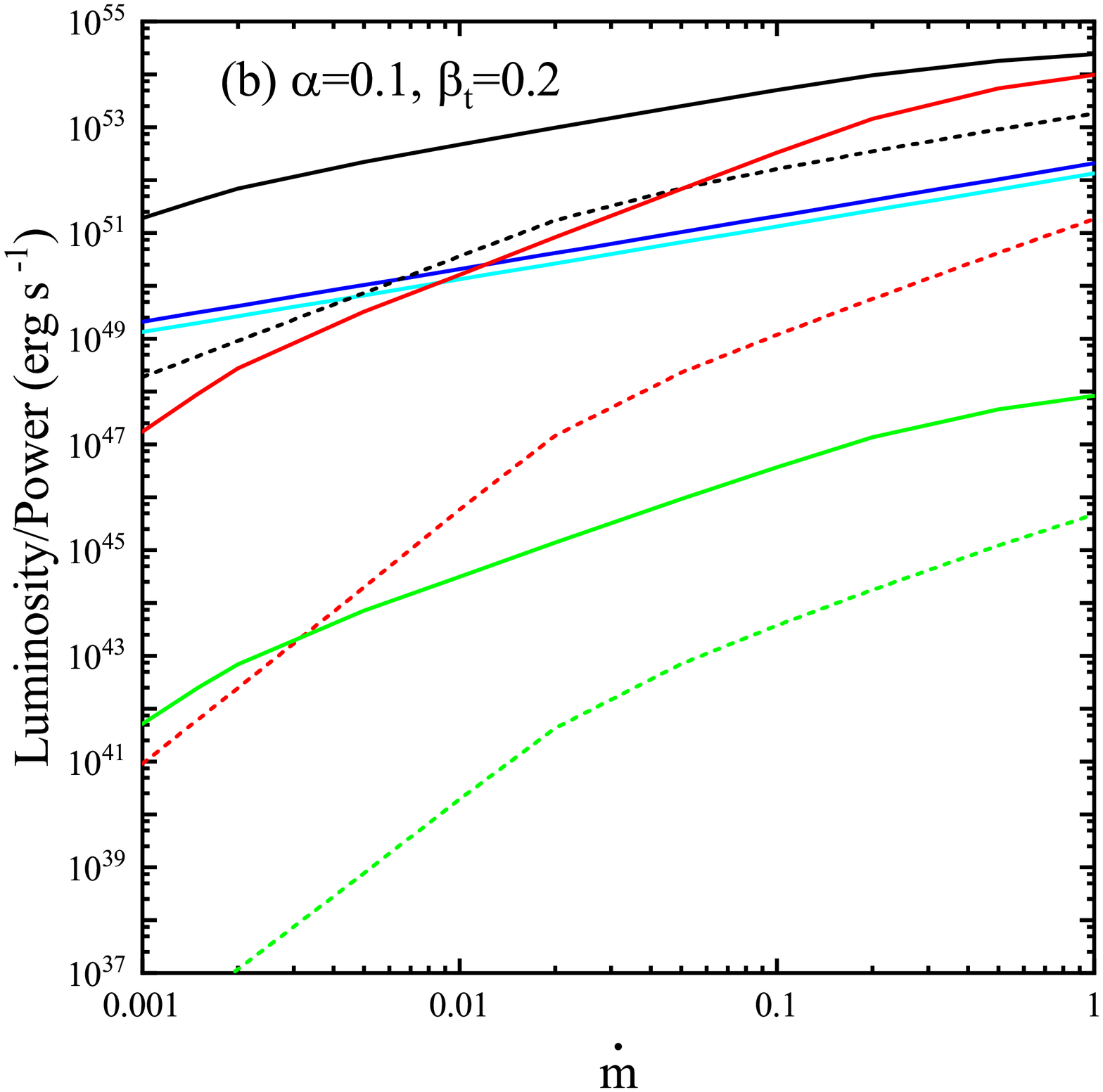}
\includegraphics[width=0.3\textwidth]{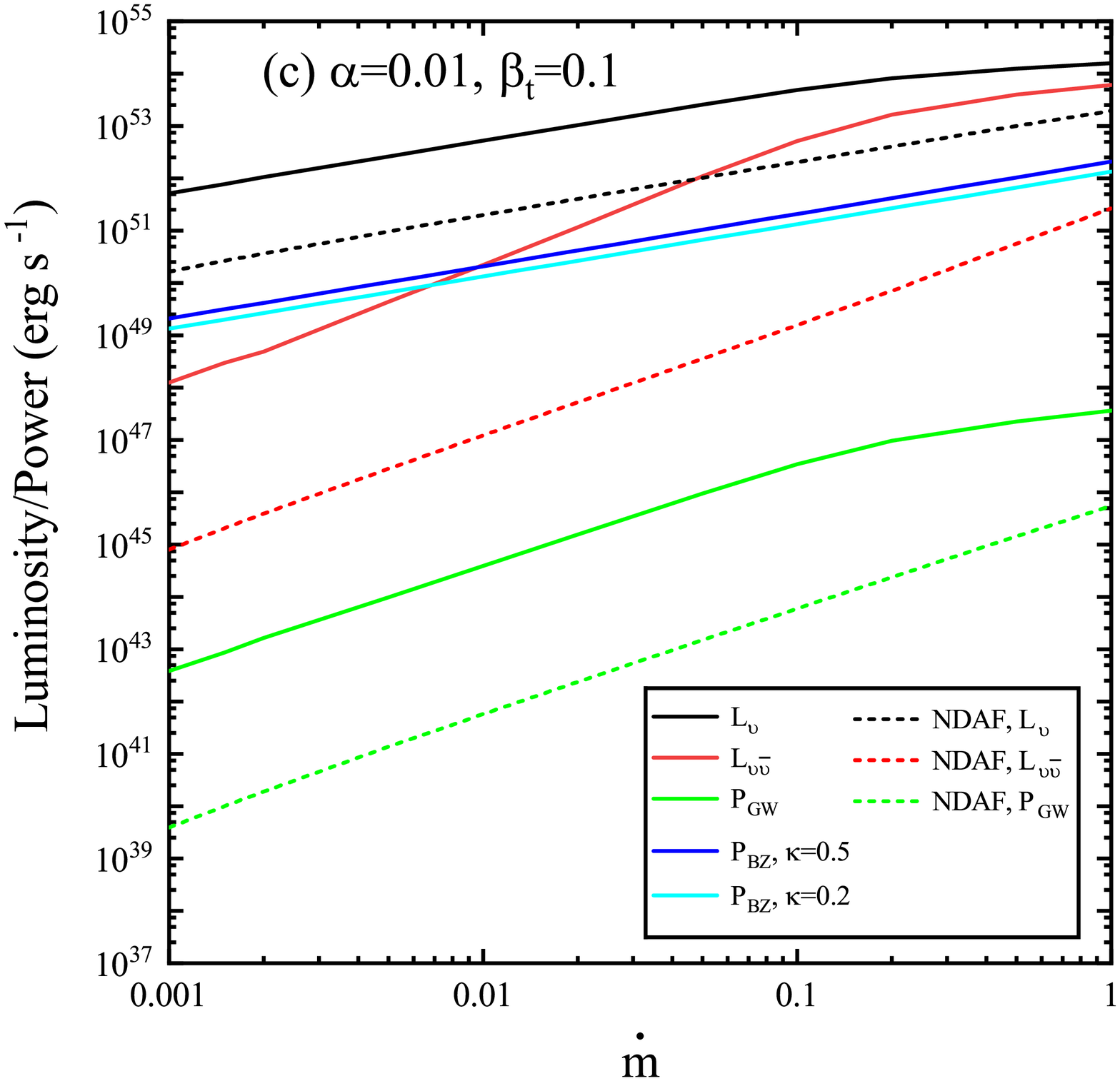}
\caption{Variations in the neutrino luminosity $L_{\nu}$, the neutrino annihilation luminosity $L_{\nu \bar{\nu}}$, the GW power $P_{\rm GW}$, and the BZ power $P_{\rm BZ}$ with the mass accretion rate for different $\alpha$ and $\beta_{t}$. Here we take $n=4$ and $a_{*}=0.8$.}
\label{fig4}
\end{figure*}

\begin{figure*}
\centering
\includegraphics[width=0.3\textwidth]{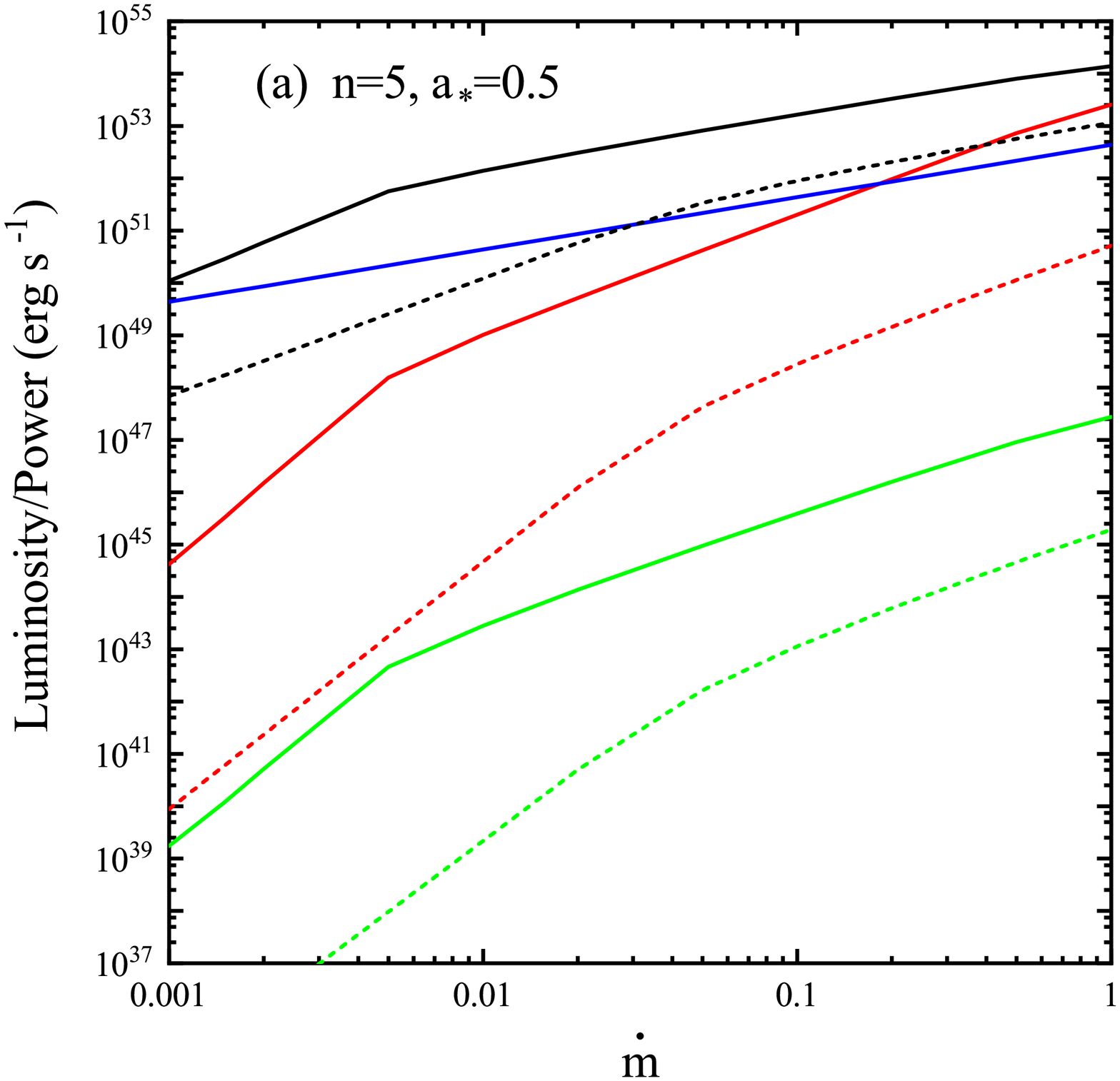}
\includegraphics[width=0.3\textwidth]{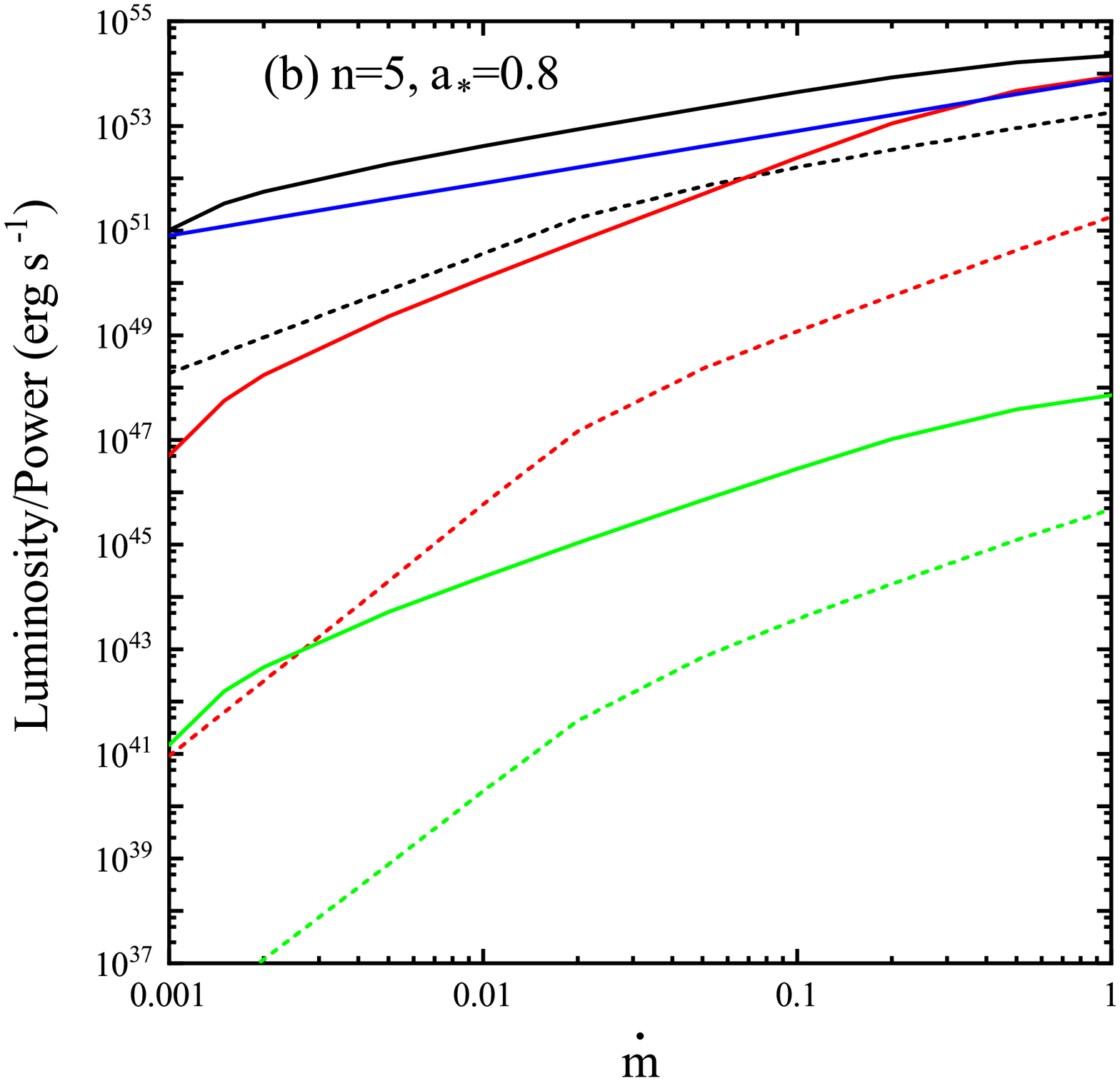}
\includegraphics[width=0.3\textwidth]{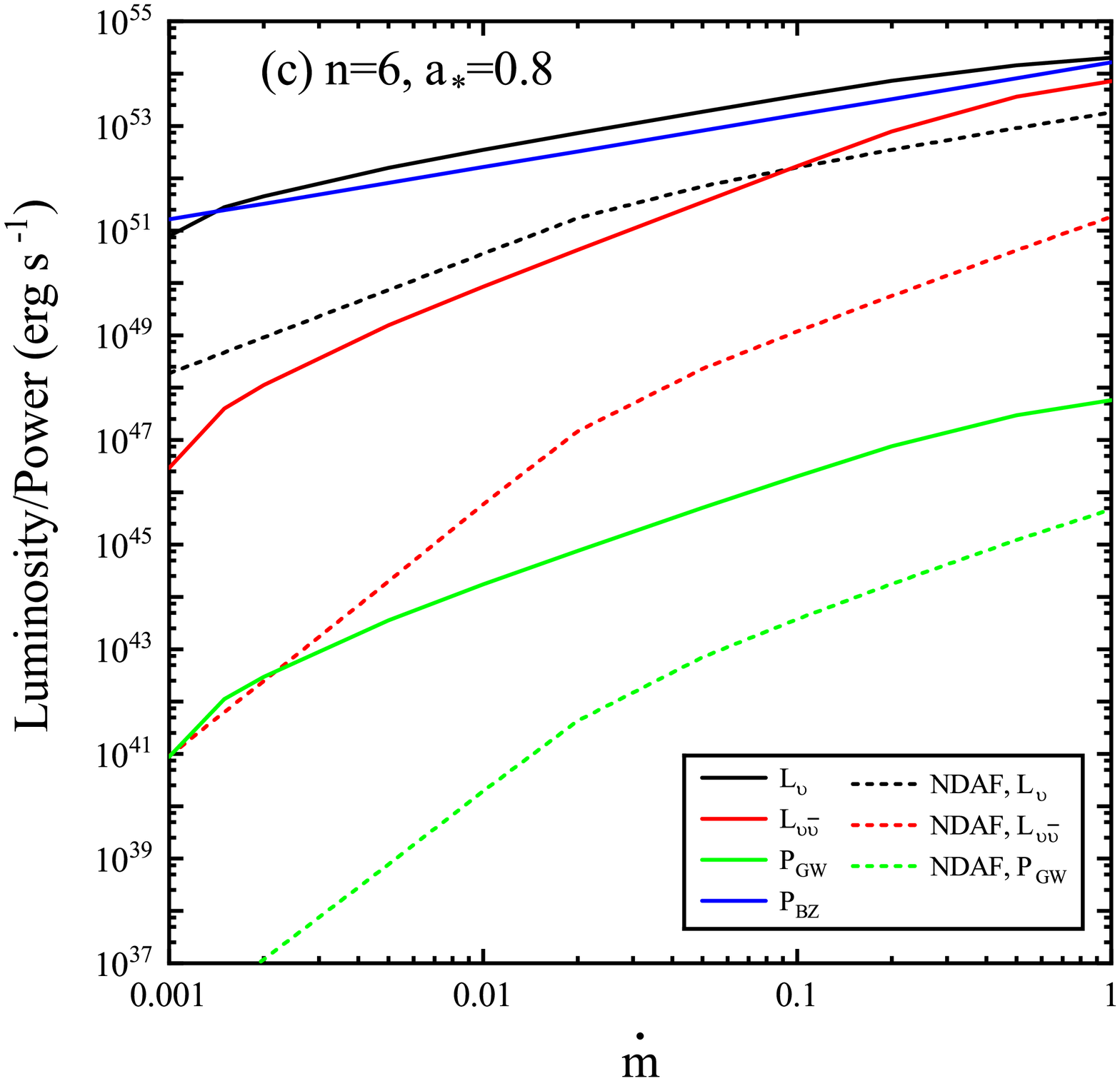}
\caption{Variations in the neutrino luminosity $L_{\nu}$, the neutrino annihilation luminosity $L_{\nu \bar{\nu}}$, the GW power $P_{\rm GW}$, and the BZ power $P_{\rm BZ}$ with the mass accretion rate for the different $n$ and $a_{*}$.  Here we take $\alpha=0.1$, $\beta_{t}=0.1$ and $\kappa =0.5$.}
\label{fig5}
\end{figure*}

For a rapidly spinning BH, the MC torque will transfer angular momentum from the BH to the inner region of the disc, resulting in energy dissipation and increasing the luminosity. For the case of BZ-MC coexistence, variations in the proportions of the two components lead to changes in the dominance of the BZ power and the neutrino annihilation luminosity. As the dominant position of the two mechanisms changes, a transition from a thermally dominated fireball to a Poynting-flux-dominated flow may be observed in some GRBs \citep[e.g.,][]{Lei2017}. \citet{Zhang2018} studied the time-resolved spectra of GRB 160625B which is composed of three sub-bursts. They found that the spectral properties of the first two sub-bursts are distinctly different, and observed the transition from thermal to non-thermal radiation between well-separated emission episodes. Such a transition indicated clearly that there is a change of jet composition from a fireball to a Poynting-flux-dominated jet.

\subsection{MeV neutrinos from MNDAFs}

During and immediately after the core-collapse of a massive star ($M \gtrsim 8 ~M_{\odot}$) at its final stages, MeV neutrinos are produced and radiated at very high rates. The supernova (SN) neutrino has been detected from SN 1987A in the Large Magellanic Cloud (LMC). However, only eleven events were detected by Kamiokande \citep{Bionta1987,Hirata1987}, whose distance, duration, and total neutrino energy are about $\rm 50 ~kpc$, $\rm 13~ s$, and $\rm 10^{53}~ ergs$, respectively. The capability of next-generation neutrino detectors has been greatly enhanced. In the near future, the Hyper-Kamiokande detector could detect 7,000-10,000 neutrino events in the case of the LMC \citep{Abe2011}. At 10 kpc, a core-collapse SN would register during the first second of post-bounce evolution $\sim 10^{4}$ times as many events in IceCube as were culled by Kamioka from SN 1987A. The other detectors, such as JUNO, Super-Kamiokande, and DUNE would witness $\sim 10^{2}$ times more events \citep[e.g.,][]{Seadrow2018}.

MeV neutrinos from NDAFs are also possibly detected by the next-generation neutrino detectors \citep[see e.g.,][]{Nagataki2002,Caballero2014,Caballero2016,Liu2016}. Based on the event rate of SNe in the local universe, \cite{Liu2016} estimated the GRB-related NDAF detection rates of Hyper-Kamiokande, JUNO, and LENA as 0.1-0.25, 0.095, and 0.095 per century, respectively. The detection rates will be slightly enhanced due to the emergence of the MC process.

\begin{figure*}
\centering
\includegraphics[width=8cm,height=8cm]{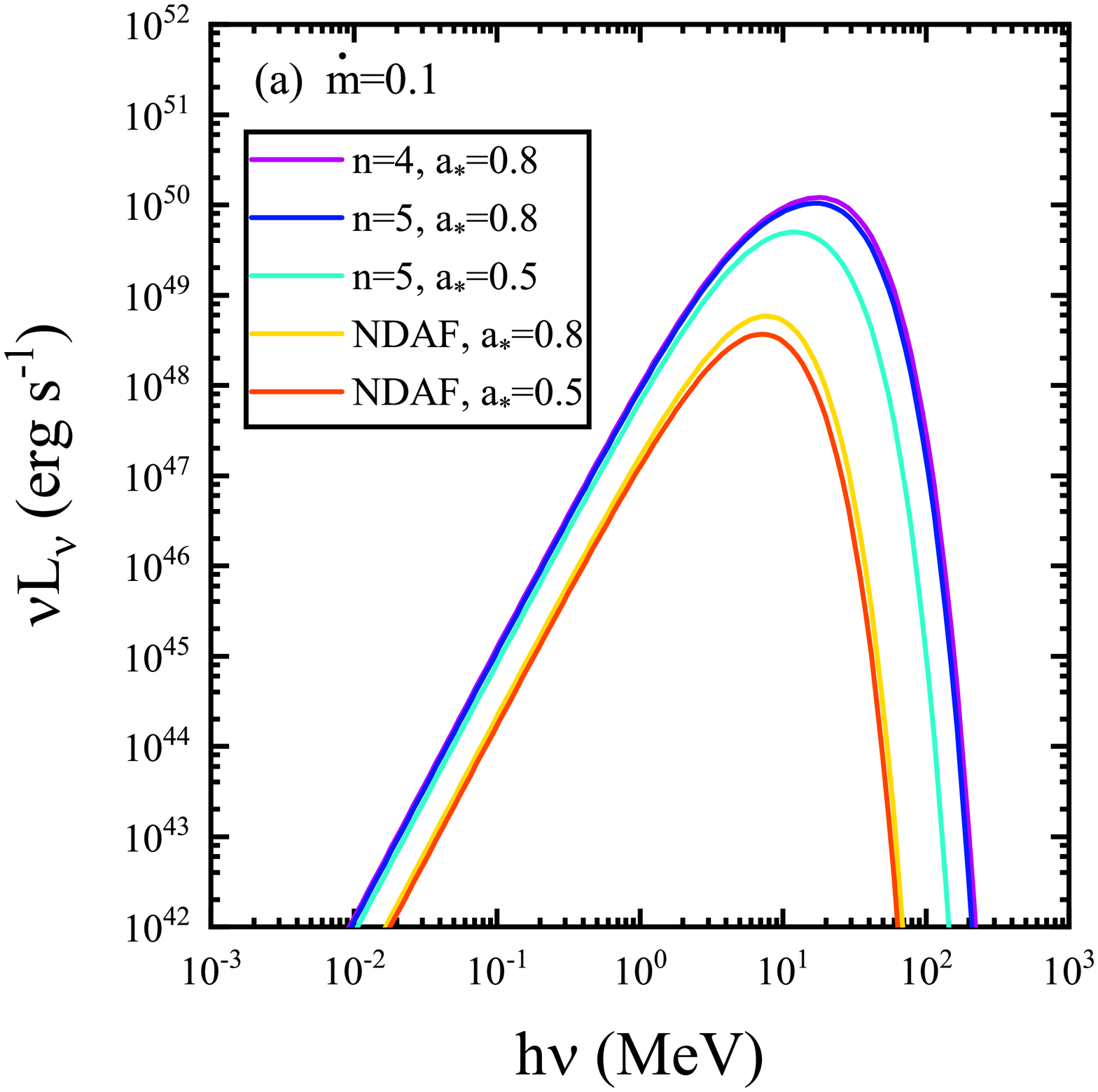}
\includegraphics[width=8cm,height=8cm]{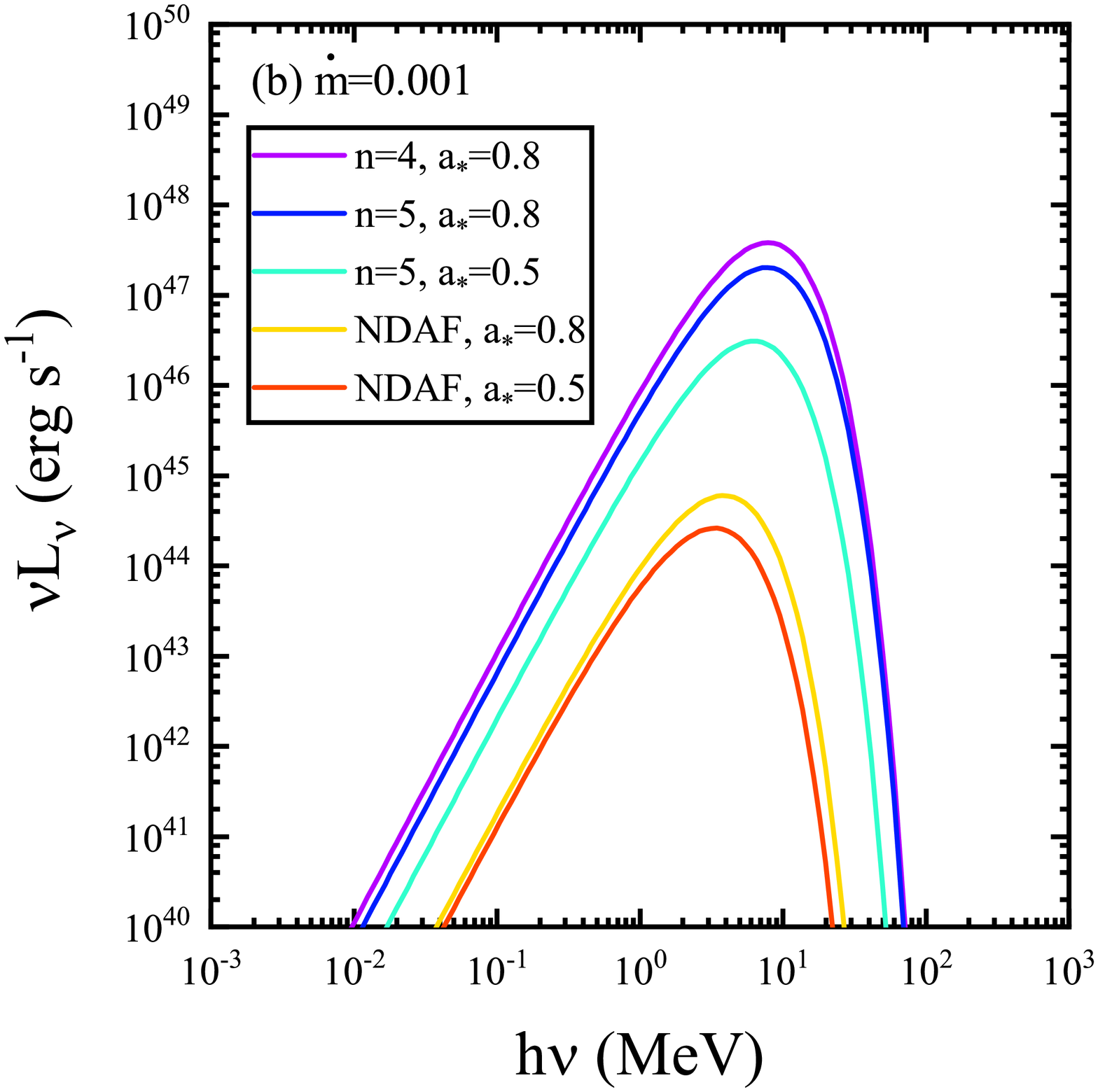}
\caption{Electron neutrino spectra of MNDAFs as the central engine of GRBs. The purple, blue, green lines indicate the electron neutrino spectra of MNDAFs with ($n$, $a_{*}$)=$(4, 0.8), (5, 0.8)$, and (5, 0.5), respectively. The electron neutrino spectra of NDAFs with $a_{*}$=0.8 and 0.5 are displayed by orange and red lines, respectively. Here, the viewing angle is adopted as $1^{\circ}$, which correspond to the case of GRBs.}
\end{figure*}

Figure 6 shows the electron neutrino spectra of MNDAFs. The purple, blue, and green lines indicate the electron neutrino spectra of MNDAFs with ($n$, $a_{*}$)=(4, 0.8), (5, 0.8), and (5, 0.5), respectively. The electron neutrino spectra of NDAFs with $a_{*}$=0.8 and 0.5 are displayed by orange and red lines, respectively. Here, the viewing angle is adopted as $1^{\circ}$, which corresponds to the case of GRBs \citep{Liu2016}.
Different $n$, $a_{*}$ and $\dot{m}$ correspond to different neutrino luminosity. Integrating neutrino luminosity over GRB duration, we can obtain the total energy $E_{\nu}$. Here we adopt $T_{\rm GRB}$ = 0.5 and 10 s as the typical duration of SGRBs and LGRBs. For MNDAF with $\dot{m}=0.1$, the $E_{\nu}$ of LGRBs and SGRBs are $\sim 10^{54} \rm~ ergs$, $\sim 10^{53} \rm ~ergs$, respectively. The $E_{\nu}$ of NDAFs with $\dot{m}=0.1$ are lower than values of MNDAF for one or two orders of magnitude. For $\dot{m}=0.1$, the peaks of the spectra of MNDAFs with ($n$, $a_{*}$) = (4, 0.8), (5, 0.8), and (5, 0.5) are about 18, 16, and 12$~\rm MeV$, while the peaks of the spectra of NDAFs with $a_{*}$ = 0.8 and 0.5 are $\sim$ 7 and 8$~\rm MeV$. The typical peak energy of neutrinos from MNDAFs is about two times higher than that from NDAFs.

Comparing these lines, we can find that the electron neutrino luminosities and the peaks of spectra increase as the increasing of BH spins and the ratios of MC magnetic fields. Comparisons between 6(a) and 6(b) indicate that the peaks of the spectra of MNDAFs are also improved by the growth of accretion rate. Similar to NDAFs, the Urca process is also the dominant neutrino cooling processes in MNDAFs, so the electron neutrinos and anti-neutrinos are the main products. There is little difference between the energy spectra of electron neutrinos and anti-neutrinos \citep[e.g.,][]{Liu2016}. Furthermore, the shapes and peak energy of neutrino spectra of MNDAFs with $\dot{M} \gtrsim 0.1 M_\odot~\rm s^{-1}$ are similar to those of O-Ne-Mg core-collapse SNe \citep[CCSNe, e.g.,][]{Cherry2012,Liu2016,Wei2019}. If MNDAFs are born in the center of progenitors of CCSNe, in the initial accretion stage, the detectabilities on MeV neutrinos from MNDAFs might approach those from CCSNe by future detectors.

It should be noticed that the accretion rate could be greater than $0.1~M_{\odot}~\rm s^{-1}$, even exceed $1~ M_{\odot} ~\rm s^{-1}$ at the early stage of massive star collapse. \cite{Wei2019} investigated the time-integrated spectra of the neutrinos from NDAFs with different masses and metallicities of progenitor stars. We found that the peak energies of spectra are $\rm 10-20~ MeV$ and a low metallicity is beneficial to the production of low-energy ($ \lesssim 1 \rm ~MeV $) neutrinos. A higher accretion rate corresponds to a higher disc temperature, which mainly produces higher-energy neutrinos. The total number of neutrino counts is proportioned to the accretion rate. The influence of the BH spin on the accretion disc neutrino detection by Super-Kamiokande has been studied by \citet{Caballero2016}. They found that the total number of counts can be about an order of magnitude higher for spinning BHs than non-rotating BHs in the case of $ \dot{M}=3~M_{\odot}~\rm s^{-1}$.

\subsection{GWs from MNDAFs}

\begin{figure*}
\centering
\includegraphics[width=5.4cm,height=5.4cm]{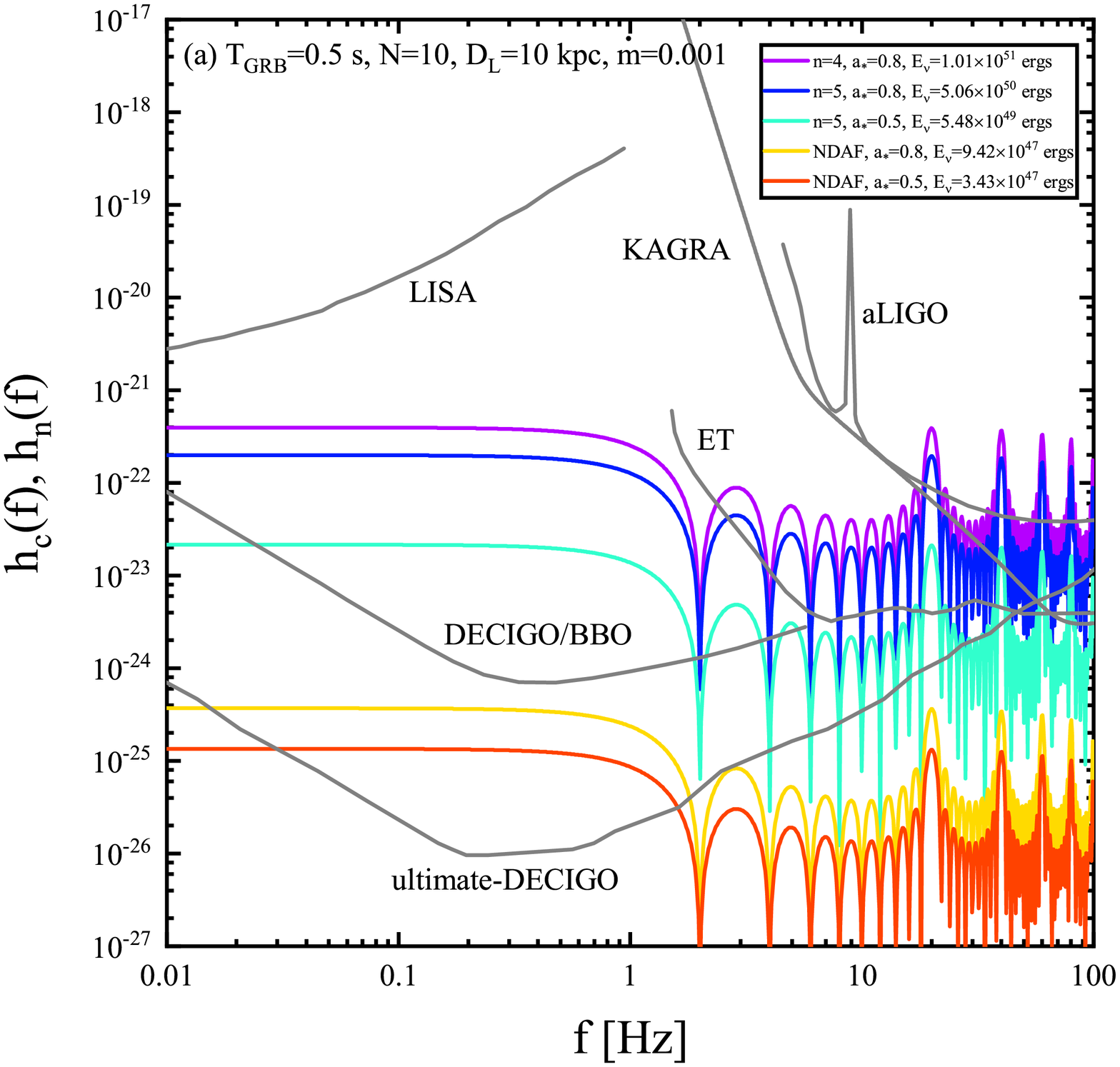}
\includegraphics[width=5.4cm,height=5.4cm]{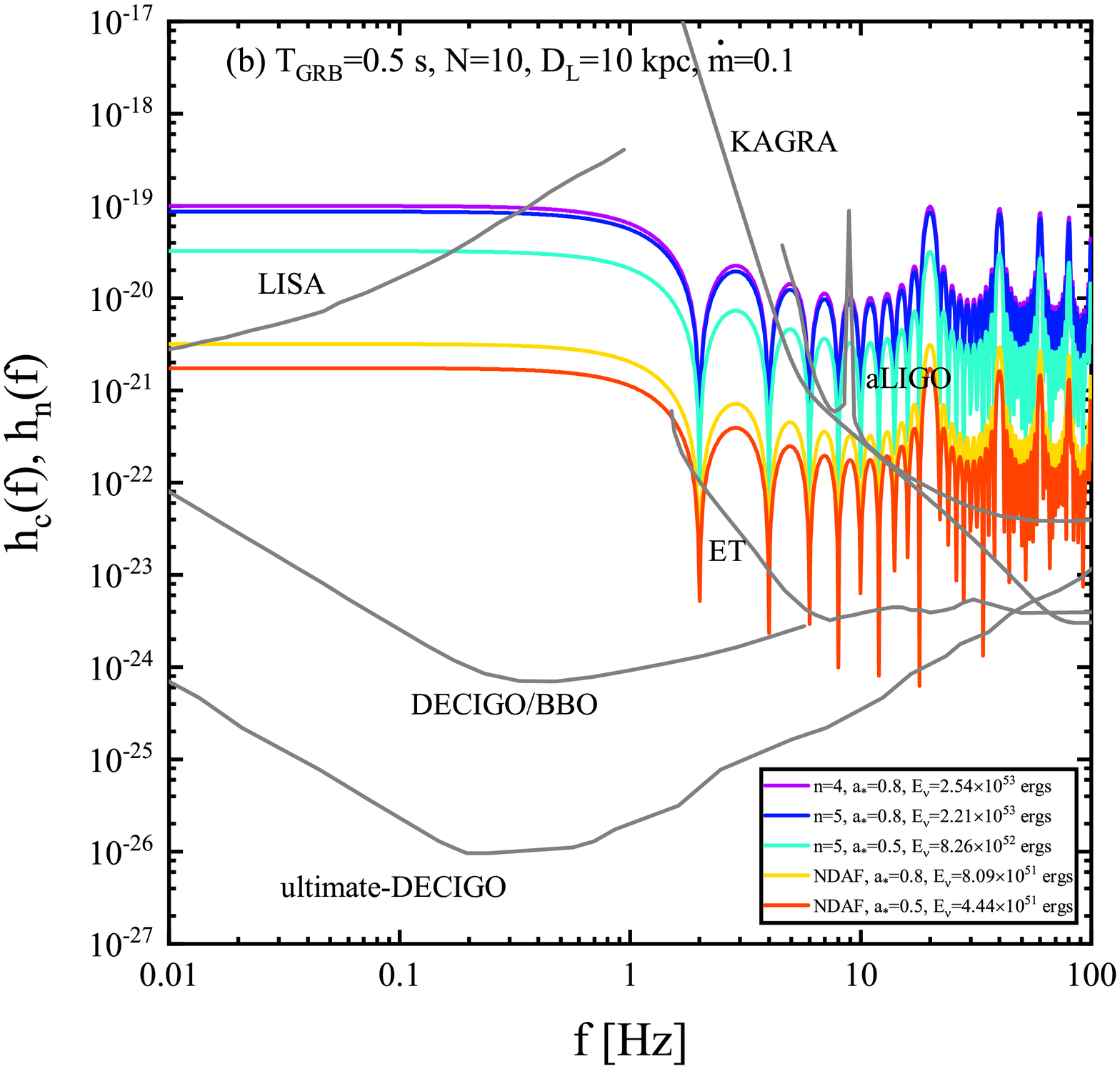}
\includegraphics[width=5.4cm,height=5.4cm]{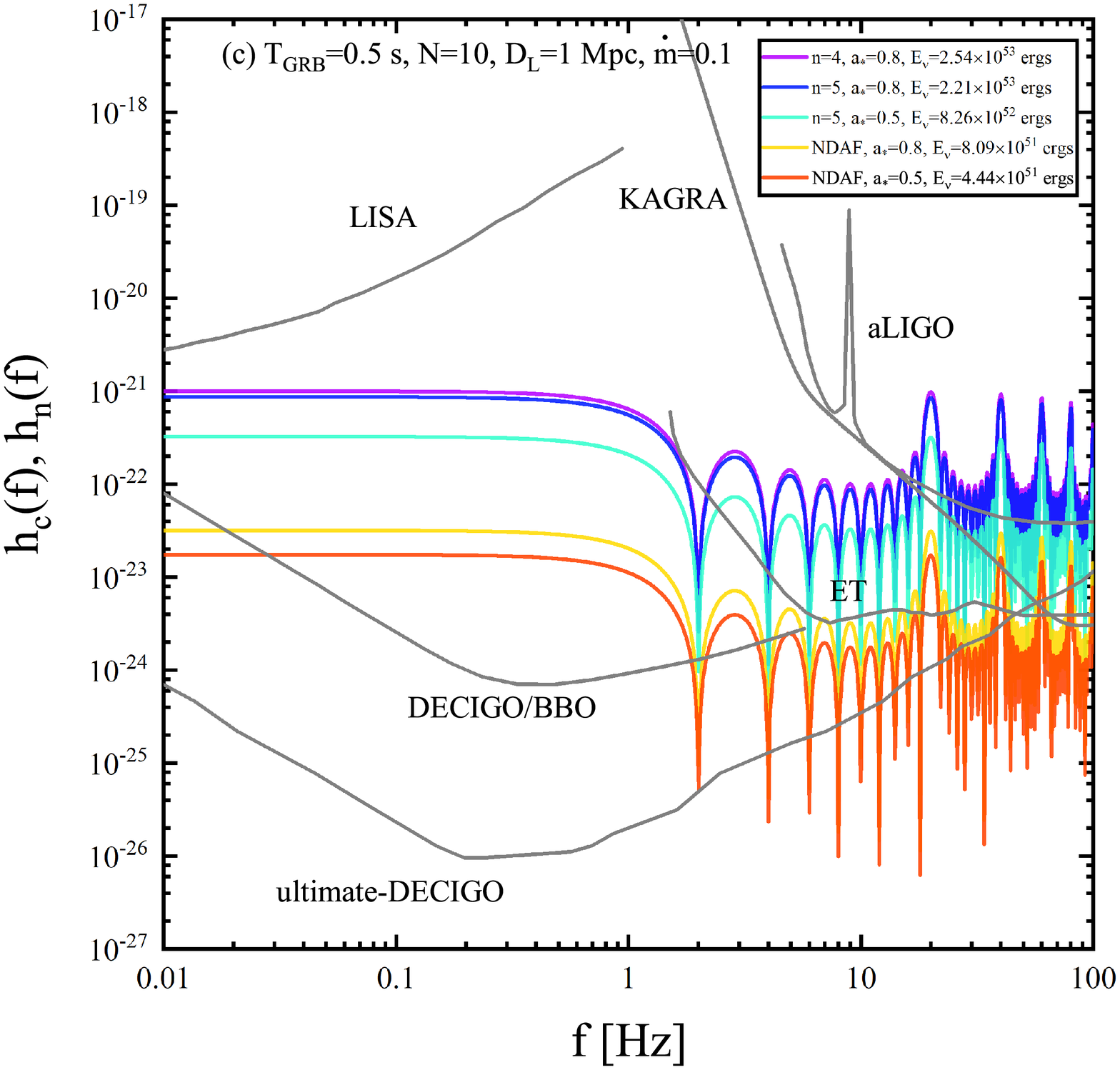}
\includegraphics[width=5.4cm,height=5.4cm]{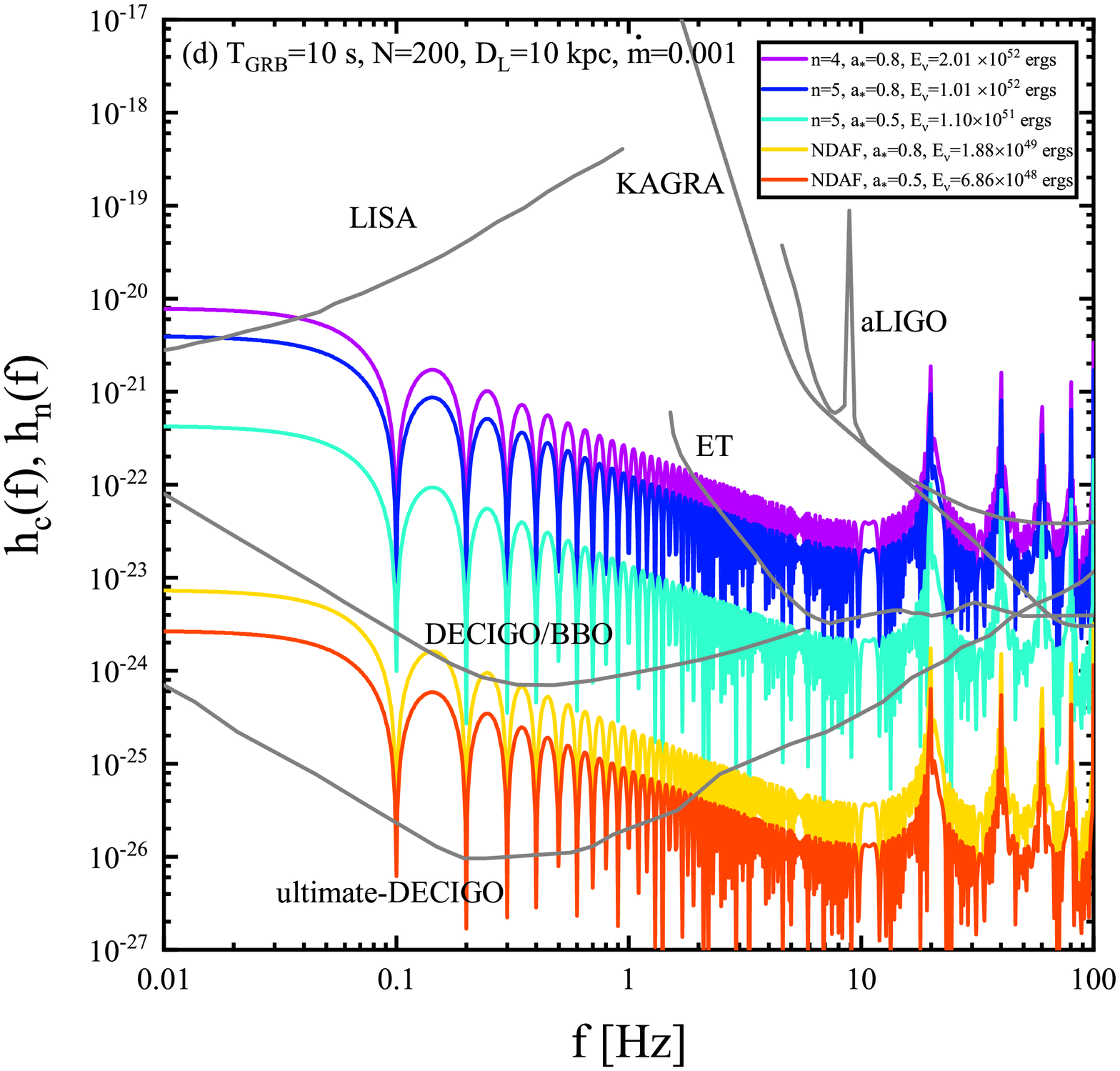}
\includegraphics[width=5.4cm,height=5.4cm]{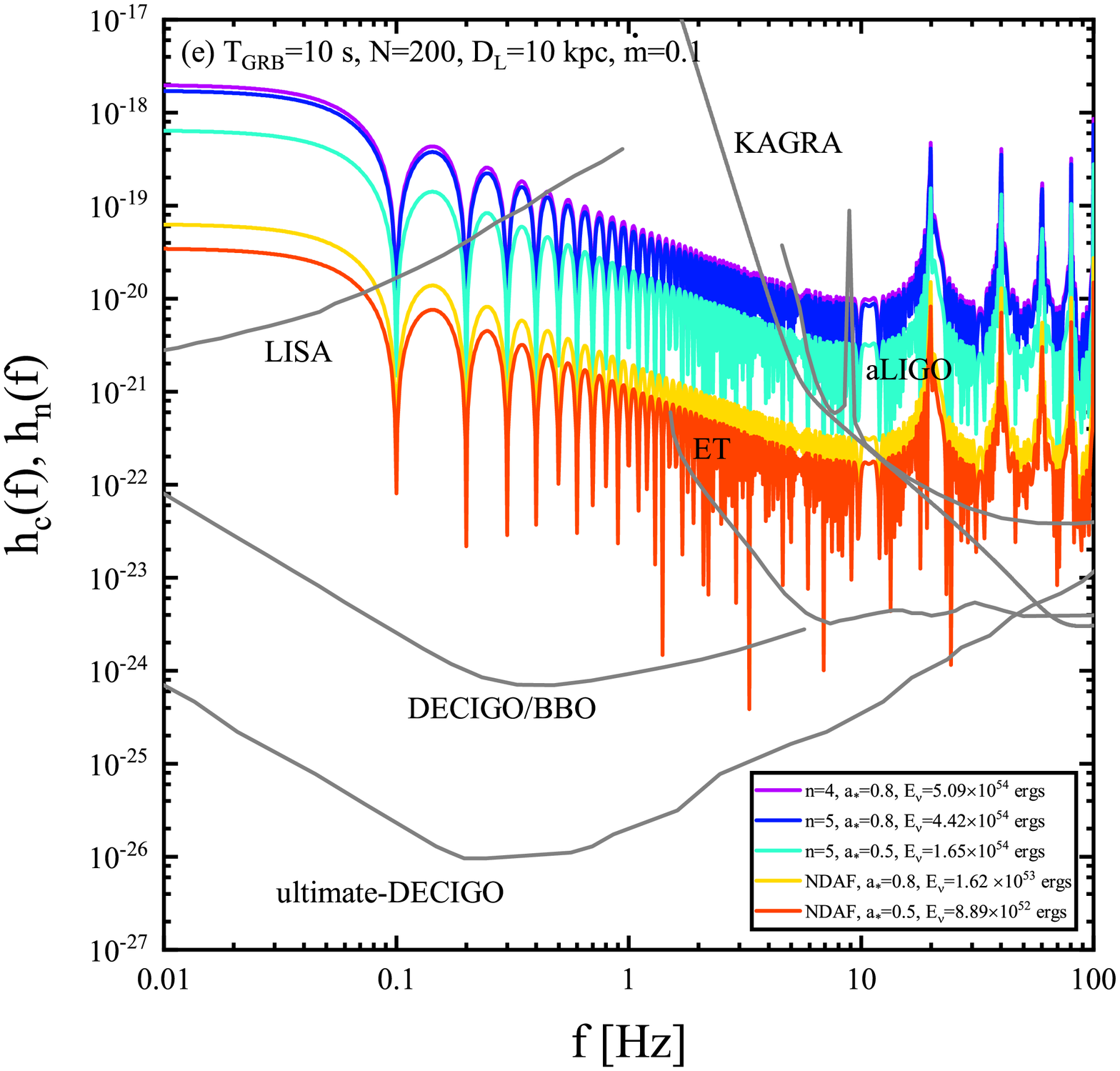}
\includegraphics[width=5.4cm,height=5.4cm]{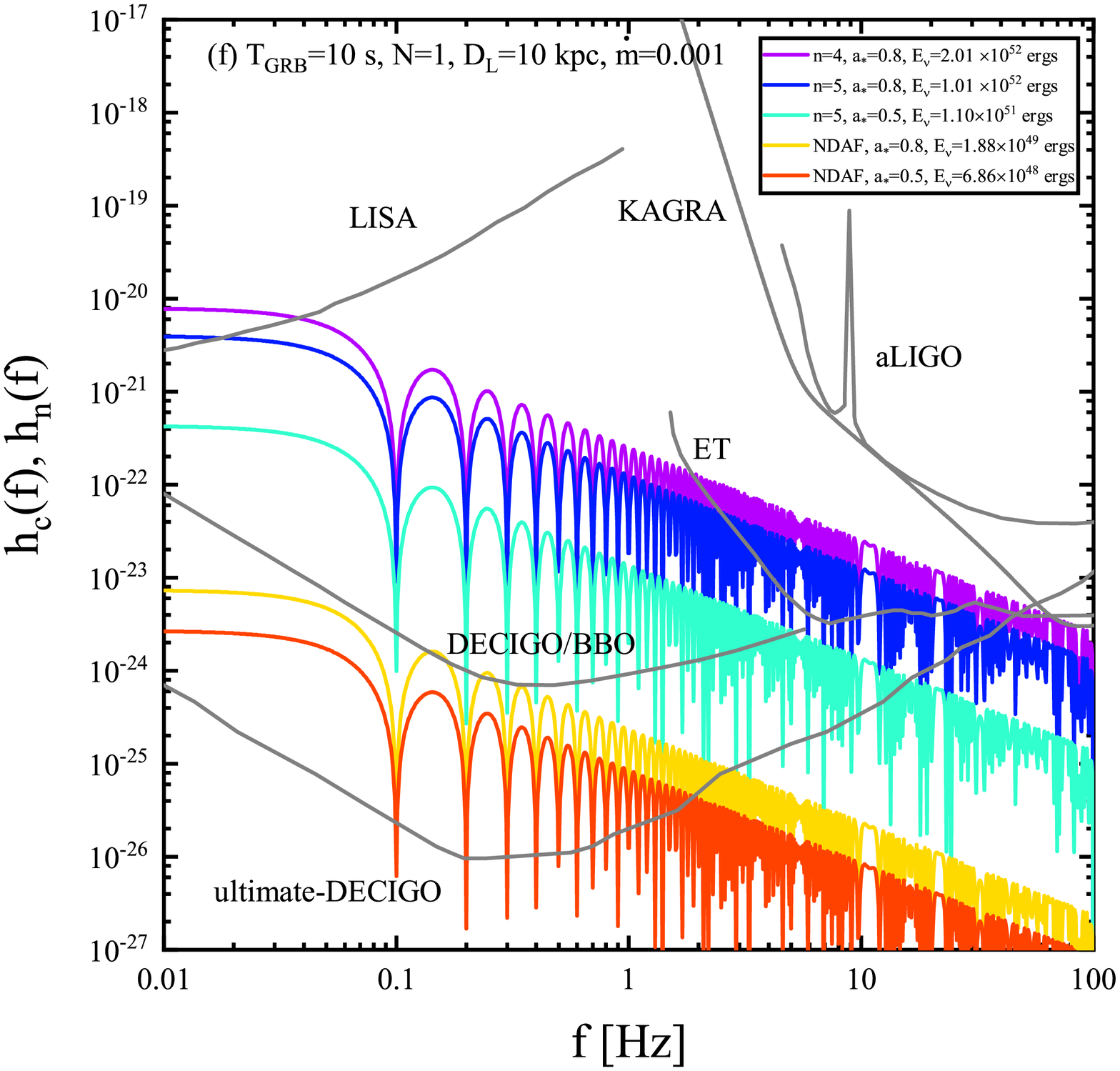}
\caption{GW strains of MNDAFs as the central engine of GRBs. The purple, blue, green lines indicate the GW strains of MNDAFs with ($n$, $a_{*}$)=(4, 0.8), (5, 0.8), and (5, 0.5), respectively. The GW strains of NDAFs with $a_{*}$=0.8 and 0.5 are displayed by orange and red lines, respectively. In all six figures, the gray lines show the sensitivity lines (the noise amplitudes $h_{n}$) of aLIGO, KAGRA, ET, LISA, DECIGO/BBO, and ultimate-DECIGO.}
\label{fig7}
\end{figure*}

To explore the detectability of GWs from MNDAFs by the detectors, we calculated GW strains as shown in Figure 7. The purple, blue, and green lines indicate the GW strains with ($n$, $a_{*}$) = (4, 0.8), (5, 0.8), and (5, 0.5), respectively. The GW strains of NDAFs with $a_{*}$ = 0.8 and 0.5 are displayed by orange and red lines, respectively. The values of $E_{\nu}$ could be found in legends. In all six figures, the gray lines show the sensitivity lines (the noise amplitudes $h_{n}$) of aLIGO, KAGRA, ET, LISA, DECIGO/BBO, and ultimate-DECIGO. One can roughly read the SNR from these figures.  Figures 7 (a)-(e) correspond to strains of GWs from MDNAFs as the central engines of GRBs with multi-pulses ($\delta t$ = 0.005 s), while Figure 7(f) is for single pulse.

It is easy to find that the GW strains of MNDAFs are several orders of magnitude stronger than those of NDAFs. We can see that the GW strains are positively correlated with the BH spins and accretion rates, and have negative correlations with the distances of the sources. As shown in Figure 7(a), the GW from MNDAF with $T=0.5$, $n=5$, $a_{*}=0.8$, and $E_{\nu}=5.06\times10^{50}\rm~ergs$ (the blue line) can be detected by ET in $~2-100 \rm ~Hz$, $\sim 20-100 \rm ~Hz$ by KAGRA and aLIGO, and $\sim 50-100 \rm~Hz$ by ultimate-DECIGO at a distance of $10\rm~ kpc$. While the the GWs from NDAFs with $T=0.5$, $a_{*}=0.8$, and $E_{\nu}=9.42\times10^{47}\rm~ ergs$ (the orange line) can be detected  by ultimate-DECIGO in the detectable frequency $\sim 0.01-2~\rm Hz$. Figures 7 (b) and 7(c) show the GW from the source with the same parameters but different $D_{\rm L}$. At the distance $\rm 1~Mpc$, KAGRA, aLIGO and ET are expected to detect the GW from MNDAFs with $E_{\nu}\sim 10^{53}\rm~ergs$. And the GW from NDAFs with $E_{\nu}\sim 10^{51}\rm~ergs$ at $1 ~\rm Mpc$ might be detected by ultimate-DECIGO in $\sim 2-100~\rm Hz$, and in the detectable frequency $\sim 0.01-6 \rm ~Hz$ by DECIGO/BBO. Furthermore, by comparing (d) and (f), we notice that for the same $T$, $E_{\nu}$, and $D_{\rm L}$, the spectra of GRBs for multiple pulses and single pulse are coincide with each other in the low frequency range but very different in the high-frequency range. This is because the multiple bursts are caused by many pulses within the short timescale, and long-term behaviours are independent from the detail of the bursts.

\section{Conclusions and Discussion}\label{Summary}	

In this paper, we have studied neutrino-cooled discs by taking into account the coexistence of the MC and BZ processes. A competitive relation exists between the MC and BZ mechanisms, which leads to different dominant approaches to release energy, i.e., neutrino annihilation or BZ jets. The luminosities of neutrino emission and annihilation should be significantly enhanced by MC effects, and the neutrino luminosity is positively correlated with the GW power. The GW power can be enhanced by the MC process by up to no more than six orders of magnitude for mass accretion rates in the range of 0.001-1 $M_{\odot}~\rm s^{-1}$. Rather, if the BZ mechanism is the dominant power source of GRBs, then the GWs originating from anisotropic neutrino emission will be severely restrained.

Observations with the $\emph{Swift}$ and Fermi satellites show that following the steep decline, many GRBs exhibit flares \citep[e.g.,][]{Burrows2005,Zhang2006,Chincarini2007,Falcone2007,Yi2016,Yi2017}, giant bumps \citep[e.g.,][]{Wu2013,Hou2014} or plateaus \citep[e.g.,][]{Troja2007,Liang2007,Corsi2009,Rowlinson2013} in their X-ray light curves. These phenomena are closely related to the late-time activities of the GRB central engines \citep[e.g.,][]{Bernardini2011,Liu2017a,Liu2018,Liu2019}. The magnetic BH-hyperaccretion system is one of the leading models for interpreting these GRB characteristics.

\cite{Luo2013} studied the possibility of a remnant disc after the prompt emission phase with the MC effect powering X-ray flares via the neutrino annihilation process. The results show that the annihilation luminosity can be significantly enhanced due to the coupling effects for the range of $\dot{m}$ from $0.001 ~M_{\odot} ~\rm s^{-1}$ to $0.1~ M_{\odot} ~\rm s^{-1}$. It denotes that a remnant disc with $\lesssim 0.5~ M_{\odot}$ may power most of the X-ray flares with a rest-frame duration of less than 100 s. However, neutrino annihilation fails to produce long-term X-ray activities in GRBs because it requires a relatively high accretion rate. The GRB average X-ray flare luminosity was found to decline with a power law in time: $L \propto t^{-2.7}$ \citep[e.g.,][]{Margutti2011}. For GRB flares with duration from several seconds to ten thousands seconds in rest frame, the luminosity from $\rm 10^{50}~ erg ~s^{-1}$ down to $\rm 10^{44}~ erg ~s^{-1}$, and the corresponding accretion rate from $10^{-3} ~M_{\odot}~\rm s^{-1}$ down to $10^{-9} ~M_{\odot} ~\rm s^{-1}$. Thus, our MNDAF model might be a plausible candidate for powering giant flares or bumps via transfer of the mechanism of releasing energy from neutrino annihilation to BZ jets, as the requirements of the BZ mechanism in terms of the accretion rate are far below those of neutrino annihilation for a certain GRB \citep[e.g.,][]{Liu2015,Liu2017a}.

Furthermore, \citet{Kumar2008a,Kumar2008b} proposed that the different segments in the light curves of long-duration GRBs correspond to the accretion of different zones in the progenitor stars. Some materials in the stellar core collapse into the BH, and the rest are accreted onto the newborn BH can produce the prompt emission. The plateaus of GRBs might originate from accretion of the relatively low-density stellar envelope. Features of the plateaus depend on the radius and mean specific angular momentum of the stellar envelope. Both the neutrino annihilation luminosity and the BZ power depend on the accretion rates in the different relations, i.e., $L_{\nu\bar{\nu}} \sim \dot{M}^{9/4}$ and $P_{\rm BZ}\sim \dot{M}$. When the accretion rate varies with time, which mechanism is dominant might determine whether a plateau exists \citep[e.g.,][]{Liu2018,Liu2019}.

It is well known that accretion discs can sustain large-scale magnetic fields in some general relativistic magnetohydrodynamic simulations \citep[e.g.,][]{De2003,McKinney2006,Dexter2010,Tchekhovskoy2010,Tchekhovskoy2011,McKinney2012,Narayan2012,Jiang2014,McKinney2014,Yuan2014,Ryan2015,White2016,Chandra2017,Chael2019,Liska2019}. Apart from the BZ and MC mechanisms, there are several other possible types of field geometries that might exist in BH accretion systems \citep[e.g.,][]{Hirose2004,McKinney2005}, including the Balbus-Hawley instability inside of the discs \citep{Balbus1991}, disc surface reconnections driven by the Parker instability, and the Blandford-Payne \citep[BP,][]{Blandford1982} process \citep[for a detailed review, see, e.g.,][]{McKinney2005, Jafari2019}. Magnetic fields play important roles in the dynamics of accretion discs \citep[e.g.,][]{Begelman1984,Punsly2001,Krolik2005}, especially in transporting the angular momentum and energy of the BHs or discs. In this paper, we focus on the BZ and MC processes and ignore other major phenomena affecting the dynamics of the disc, such as the role of turbulence and magnetic stochasticity \citep[e.g.,][]{Eyink2013,JafariVishniac2019}, the role of reconnection, magnetic buoyancy, and turbulent pumping \citep[e.g.,][]{Jafari2018} and flux expulsions \citep[e.g.,][]{Spruit2005,Guan2009}.

MC geometries appear rarely in some numerical calculations and simulations \citep[e.g.,][]{Balbus1998,Hawley2011,Cao2014,Cao2019}. A possible reason might be that most of the initial large-scale fields of the disc are open magnetic lines, which are difficult to be squeezed because the large loops are tied onto the disc. Actually, magnetic buoyancy results in the formation of a flux rope out of the disc surface, and magnetic reconnection might power episodic jets \citep{Yuan2012}. When these field lines can continuously move to the BH via the accretion process and one of the magnetic footpoints falls into the ergosphere, temporary MC geometries may be formed. If a BH is surrounded by a face-on disc with such temporary field geometries, then several rotating hot spots could be observable. This model might be used to explain the observations of the compact polarized `hot spot' near the last stable circular orbit of the massive BH SgrA* \citep{Abuter2018}. We will simulate the above process in future works.

\section*{Acknowledgements}
We thank Prof. Ding-Xiong Wang, Prof. Alexander Heger, and Dr. Bernhard M{\"u}ller for helpful discussion and an anonymous referee for suggestions and comments. This work was supported by the National Natural Science Foundation of China under grant 11822304. C.-Y. Song acknowledges financial support from the China Scholarship Council to work at Monash University.

\label{lastpage}
\end{document}